\documentclass[aps,prc,twocolumn,superscriptaddress,nofootinbib]{revtex4}
\usepackage{graphicx,amsfonts,amsmath,color,epsfig,epstopdf}
\usepackage{multirow}
\usepackage[normalem]{ulem}
\newcommand{\SU}[1]{\ensuremath{\mathrm{SU}( #1 )}}

\newcommand{\SpR}[1]{\ensuremath{\mathrm{Sp}( #1,\mathbb{R} )}}
\newcommand{\ME}[3]{\ensuremath{\langle #1 | #2 | #3 \rangle}}

\newcommand{\betb}{\begin{tabular}{p{4.0cm}p{9.0cm}}}
\newcommand{\entb}{\end{tabular}}

\newcommand{\ho}{\ensuremath{\hbar\Omega}}
\newcommand{\ph}[1]{\ensuremath{#1}p-\ensuremath{#1}h}
\begin{document}

\title{Understanding emergent collectivity and clustering  in nuclei \\ from a symmetry-based no-core shell-model perspective}
\author{A. C. Dreyfuss}
\affiliation{Department of Physics and Astronomy, Louisiana State University, Baton Rouge, LA 70803, USA}

\author{K. D. Launey}
\affiliation{Department of Physics and Astronomy, Louisiana State University, Baton Rouge, LA 70803, USA}

\author{T. Dytrych}
\affiliation{Department of Physics and Astronomy, Louisiana State University, Baton Rouge, LA 70803, USA}
\affiliation{Nuclear Physics Institute, Academy of Sciences of the Czech Republic, 250 68 \v{R}e\v{z}, Czech Republic}

\author{J. P. Draayer}
\affiliation{Department of Physics and Astronomy, Louisiana State University, Baton Rouge, LA 70803, USA}

\author{R. B. Baker}
\affiliation{Department of Physics and Astronomy, Louisiana State University, Baton Rouge, LA 70803, USA}

\author{C. M. Deibel}
\affiliation{Department of Physics and Astronomy, Louisiana State University, Baton Rouge, LA 70803, USA}

\author{C. Bahri}
\affiliation{Department of Physics and Astronomy, Louisiana State University, Baton Rouge, LA 70803, USA}

\begin{abstract}

We present a detailed discussion of the structure of the low-lying positive-parity energy spectrum of $^{12}$C from a no-core shell-model perspective. The approach utilizes a fraction of the usual shell-model space and extends its multi-shell reach via the symmetry-based no-core symplectic shell model (NCSpM) with a simple, physically-informed effective interaction. We focus on the ground-state rotational band, the Hoyle state and its $2^+$ and $4^+$ excitations, as well as the giant monopole $0^+$ resonance, which is a vibrational breathing mode of the ground state. This, in turn, allows us to address the open question about the structure of the Hoyle state and its rotational band. In particular, we find that the Hoyle state is best described through deformed prolate collective modes rather than vibrational modes, while we show that the higher-lying giant monopole $0^+$ resonance resembles the oblate deformation of the $^{12}$C ground state. In addition, we identify the giant monopole $0^+$ and quadrupole  $2^+$ resonances of selected light and intermediate-mass nuclei, along with other observables of $^{12}$C, including matter rms radii,  electric quadrupole moments, as well as $E2$ and $E0$ transition rates.

\end{abstract}


\maketitle

\section{Introduction}

Recent advances in nuclear modeling, greatly aided by the availability of high-performance-computing (HPC) facilities, have enabled the re-examination of a long-standing challenge in nuclear physics \cite{EllisE70,EngelandE72,UegakiOAT77,Kamimura81,SuzukiH86}, namely, understanding $\alpha$-clustering and  highly-deformed spatial configurations from a microscopic many-particle perspective. These phenomena are exemplified by the elusive 7.65-MeV second $0_2^+$ (Hoyle) state of $^{12}$C \cite{B2FH} and its associated rotational excitations, which continue to be studied theoretically (e.g., \cite{Kanada98,FunakiTHSR03,YamadaS05,ChernykhFNNR07, UmarMIO10,KhoaCK11,NeffF14}, with recent reviews \cite{HoriuchiIK12,FunakiHT15}), as well as experimentally (e.g., \cite{Freer0709,Hyldegaard10,Itoh11,ZimmermanDFGS11,Zimmerman13, NPatel13,Marin14,Freer2010}). Key features of the Hoyle state have been recently revealed within the \emph{ab initio} frameworks of lattice effective field theory (EFT) \cite{EpelbaumKLM11,EpelbaumKLLM12} and Green's function Monte Carlo \cite{Wiringa12}, together with a first fully microscopic no-core symplectic shell-model (NCSpM) study \cite{DreyfussLDDB13}. This interest is motivated, in part, because various phenomena of astrophysical significance, such as nucleosynthesis, the evolution of primordial stars in the Universe, X-ray bursts, and core-collapse supernovae simulations, are greatly influenced by several important low-lying states in $^{12}$C \cite{Fynbo05}, including the Hoyle state and its long-debated $2^+$ and $4^+$  excitations.

In this paper, we address -- from a no-core shell-model perspective -- open questions about the structure, radii, and deformation of the Hoyle-state rotational band in $^{12}$C, and investigate the underpinning mechanism of these excitations: whether they are vibrational or shape-coexistence modes. While low-lying $0^+$ states have commonly and historically been understood as vibrational modes, a different mechanism -- shape coexistence -- has been proposed for the Hoyle-like $0^+_2$ state in $^{16}$O  \cite{RoweTW06}, and has been found to occur in many nuclei across the nuclear chart (see Fig. 8 of Ref. \cite{HeydeW11}), including $^{16}$O, $^{40}$Ca, $^{56}$Ni, and the vicinity of (single-) closed shell nuclei, together with regions around  $^{74}$Se and $^{100}$Zr, as well as $^{152}$Sm and $^{154}$Gd, up to heavy isotopes of Au, Hg, Tl, Pb, Bi, Po, and many others \cite{Kulp08,RoweW2010book,HeydeW11}. This corroborates our earlier findings of a large prolate deformation for the Hoyle state \cite{DreyfussLDDB13} that is very different from the oblate shape of the ground state of $^{12}$C. In the present study, we examine largely deformed prolate configurations favored in the low-lying energy spectrum of $^{12}$C, and the vibrational breathing mode of the oblate  ground state within a unified shell-model approach. In particular, we employ the NCSpM with a simple but physically-informed, schematic, many-nucleon interaction. The model has recently provided, with no parameter adjustments, no-core shell-model descriptions of low-lying states in deformed $sd$-shell nuclei and of phenomena tied to collectivity and alpha-clustering in $^{8}$Be \cite{LauneyDDTFLDMVB13,TobinFLDDB14}.  Here, we focus on the ground-state rotational band, the Hoyle state and its $2^+$ and $4^+$ excitations, together with the giant monopole $0^+$ resonance (GMR)  and  giant  quadrupole $2^+$ resonance (GQR) in $^{12}$C. With the aim to gain further insight into the collective and cluster-like substructures of $^{12}$C, we consider excitation energies, matter rms radii, electric quadrupole moments, $E2$ and $E0$ transition rates, and probability distributions.

We also identify giant monopole and quadrupole resonances in other $p$- and $sd$-shell nuclei, including $^{16,20}$O, $^{20,22}$Ne, and $^{20,22}$Mg. Giant multipole resonances in nuclei, such as the GMR and GQR, provide important information about nuclear structure, including information about the compressibility of nuclear matter in the case of the GMR \cite{Walecka1962,UWerntz1966}, which is often referred to as a vibrational breathing mode. Further, GMRs figure prominently in isoscalar monopole strengths \cite{Suzuki87,SuzukiH89} with a  renewed interest \cite{YamadaFMHIRST12} following the reach of experimental data  to higher excitation energies. For the lightest nuclei, the first isoscalar monopole excitations  of $^4$He has been examined within an {\it ab initio} framework \cite{BaccaBLO13,BaccaBLO15}, while the symplectic shell model of Rosensteel and Rowe \cite{RosensteelR77, Rowe85}, which adopts a symplectic \SpR{3} basis and underpins the present NCSpM, has provided successful microscopic descriptions of giant resonances of $sd$-shell and heavy nuclei \cite{HechtB82,VassanjiR83,Suzuki87,RBVArickx1990, BahriDCR90,CarvalhoR92,BahriR00}, along with $^8$Be \cite{CarvalhoRKB02}. The NCSpM is hence well-suited to identify and study giant resonances: in the model, the symplectic symmetry provides a classification of the complete translationally invariant shell-model space, while dividing the space into vertical symplectic cones;  basis states of a symplectic cone are built by the monopole and quadrupole moment operators, which describe one-phonon excitations of a giant monopole and giant quadrupole type \cite{Rowe85}. Indeed, the dominant role of the symplectic symmetry in light nuclear systems has been recently confirmed by {\it ab initio} studies of nuclei from $^6$Li to $^{16}$O \cite{DytrychSBDV_PRL07, DytrychSDBV08_review, DytrychSBDV_PRCa07,DytrychLMCDVL_PRL12}.

The paper is organized as follows. We start with a brief outline of the NCSpM model (Sec. \ref{NCSpM}), together with the model spaces and the schematic long-range interaction that are used, and show they yield results that closely agree with feasible {\it ab initio} outcomes for the ground-state rotational band of $^{12}$C. In Sec. \ref{Hoyle}, we present NCSpM results in down-selected ultra-large shell-model spaces for the Hoyle-state rotational band and its structure (including deformation, radii, quadrupole moments, and transition rates), and examine the dependence of the results on model parameters and on the size of the model space. The last section, Sec. \ref{GR}, focuses on the structure and observables for the GMR and GQR in  $^{12}$C and other selected nuclei, with a  discussion on the role of the vibrational breathing mode in the $^{12}$C excitation spectrum.

\section{No-core symplectic shell model with \SpR{3} symmetry \label{NCSpM}}
The NCSpM, as outlined in Ref. \cite{DreyfussLDDB13,TobinFLDDB14}, is based on the physically relevant symplectic \SpR{3} group \cite{RosensteelR77,Rowe85} and its embedded \SU{3} subgroup \cite{Elliott58,Elliott58b,ElliottH62}. These symmetries provide an organization of the model space into symplectic basis states (or vertical cones), as described below, which are comprised of states of definite deformation and are related via a unitary transformation to three-dimensional harmonic oscillator (HO) many-body basis states \cite{DytrychSDBV08_review}, such as the \emph{m}-scheme basis used in the no-core shell model (NCSM) \cite{NavratilVB00,BarrettNV13}.  In fact, the NCSpM and NCSM coincide for the same $N_{\max}$, where $N_{\max}$ describes the cutoff in total oscillator quanta above the lowest HO configuration for the system.

\subsection{Model space selection}
Each basis state of a symplectic \SpR{3} irreducible representation (irrep) is labeled according to the group chain \cite{Rowe85},
\begin{center}
\begin{equation}
\begin{tabular}{ccccccc}
Sp(3,$\mathbb{R}$)&$\supset$&U(3)&$\supset$&SO(3)& $\supset$ & SO(2) \\
$\sigma$ & $n\rho$ & $\omega$ & $\kappa$ & $L$ && $M$ \\
\end{tabular}
\end{equation}
\end{center}
and is constructed using the following relation with symmetrically coupled polynomials in the symplectic raising operators, $A^{(20)}$:
\begin{eqnarray}
&& | \sigma =N_{\sigma} (\lambda_{\sigma}\,\mu_{\sigma}), n=N_{n}(\lambda_{n}\,\mu_{n}),\rho, \omega =N_{\omega}(\lambda_{\omega}\, \mu_{\omega}),\kappa LM\rangle 
\nonumber \\
&&= \left[  \left[ \underbrace{A^{(20)}\times \dots \times A^{(20)}} \right]^{(\lambda_{n}\,\mu_{n})} \times 
|N_{\sigma}(\lambda_{\sigma} \mu_{\sigma})\rangle  \right]^{\rho(\lambda_{\omega}\, \mu_{\omega})}_{\kappa LM},  \nonumber \\
&& \hspace{0.65in}    N_n/2
\end{eqnarray}
where $N_{\omega}=N_{\sigma}+N_n$ is the total number of oscillator quanta ($\rho$ and $\kappa$ are multiplicity labels).  The $A^{(20)}$ operator induces  2\ho~ $1$-particle-$1$-hole (\ph{1}) monopole or quadrupole excitations (one particle raised by two shells)  together with a smaller 2\ho~\ph{2} correction for eliminating the spurious center-of-mass (CM) motion. The symplectic bandhead, $|N_{\sigma}(\lambda_{\sigma} \mu_{\sigma})\rangle$, is  the lowest-weight \SpR{3} state, which is defined by the usual requirement that the symplectic lowering operators $(A^{(20)})^\dagger$ annihilate it.  The  bandhead is an \SU{3}-coupled many-body state with a given nucleon distribution over the HO shells (that is, a set of $\{ \eta_1,\dots,\eta_A\}$ configurations with $\eta_i$ the oscillator number of the $i$-th particle for a nuclear mass number $A$). The corresponding  $N_\sigma\ho$ energy of HO quanta\footnote{This energy includes the HO zero-point energy. To eliminate the spurious CM motion, we use $N_\sigma$, for which $3/2$ is subtracted from the total HO quanta, together with symplectic generators constructed in relative coordinates with respect to the  CM. These generators are used to build the basis, the interaction, the many-particle kinetic energy operator, as well as to evaluate observables.}, together with the bandhead deformation, $(\lambda_{\sigma} \mu_{\sigma})$, serve to label the symplectic irrep. An example is shown in Table \ref{sp04} for the basis states of a $0\ho(0\,4)$ symplectic irrep up through $N_n=6$. 

Including the spin degrees of freedom requires the straightforward generalization, $|\sigma n\rho\omega\kappa (LS_{\sigma})JM_{J}\rangle=\sum_{MM_{S}}\langle LM;S_{\sigma}M_{S}|JM_{J}\rangle |\sigma n\rho\omega\kappa LMS_{\sigma}M_S\rangle$. All of the states within a symplectic irrep share the same spin value, given by the spin $S_{\sigma}$ of the bandhead $|\sigma;S_{\sigma}\rangle$. With the inclusion of the additional $\{\alpha\}$ quantum numbers to distinguish between physically distinct bandheads with the same $N_{\sigma}(\lambda_{\sigma} \mu_{\sigma})$, $|\{\alpha\}~\sigma\rangle$, the symplectic basis states span the entire shell-model space.
\begin{table}[]
\caption{Basis states of the $^{12}$C symplectic irrep $24.5(0\, 4)$, or equally $0\ho(0\,4)$, up through $N_{\max}=6$ for $L=2$ (and $M=0$, with $N_{\omega}=24.5+N_n$).}
\begin{center}
\begin{tabular}{ccccc | ccccc}
$N_{n}$	&	$(\lambda_{n}\,\mu_{n})$		&	$\rho$	&	$(\lambda_{\omega} \, \mu_{\omega})$	&	$\kappa$ &
$N_{n}$	&	$(\lambda_{n}\,\mu_{n})$		&	$\rho$	&	$(\lambda_{\omega} \, \mu_{\omega})$	&	$\kappa$
\\ \hline \hline
0	&	$(	0	\,	0	)$	&	1	&	$(	0	\,	4	)$	&	1	&	6	&	$(	6	\,	0	)$	&	1	&	$(	5	\,	3	)$	&	1	\\
2	&	$(	2	\,	0	)$	&	1	&	$(	2	\,	4	)$	&	1	&	6	&	$(	6	\,	0	)$	&	1	&	$(	4	\,	2	)$	&	1	\\
2	&	$(	2	\,	0	)$	&	1	&	$(	2	\,	4	)$	&	2	&	6	&	$(	6	\,	0	)$	&	1	&	$(	4	\,	2	)$	&	2	\\
2	&	$(	2	\,	0	)$	&	1	&	$(	1	\,	3	)$	&	1	&	6	&	$(	2	\,	2	)$	&	1	&	$(	4	\,	2	)$	&	1	\\
2	&	$(	2	\,	0	)$	&	1	&	$(	0	\,	2	)$	&	1	&	6	&	$(	2	\,	2	)$	&	1	&	$(	4	\,	2	)$	&	2	\\
4	&	$(	4	\,	0	)$	&	1	&	$(	4	\,	4	)$	&	1	&	6	&	$(	2	\,	2	)$	&	1	&	$(	3	\,	4	)$	&	1	\\
4	&	$(	4	\,	0	)$	&	1	&	$(	4	\,	4	)$	&	2	&	6	&	$(	6	\,	0	)$	&	1	&	$(	3	\,	1	)$	&	1	\\
4	&	$(	4	\,	0	)$	&	1	&	$(	3	\,	3	)$	&	1	&	6	&	$(	2	\,	2	)$	&	1	&	$(	3	\,	1	)$	&	1	\\
4	&	$(	4	\,	0	)$	&	1	&	$(	2	\,	2	)$	&	1	&	6	&	$(	2	\,	2	)$	&	1	&	$(	2	\,	6	)$	&	1	\\
4	&	$(	4	\,	0	)$	&	1	&	$(	2	\,	2	)$	&	2	&	6	&	$(	2	\,	2	)$	&	1	&	$(	2	\,	6	)$	&	2	\\
4	&	$(	0	\,	2	)$	&	1	&	$(	2	\,	2	)$	&	1	&	6	&	$(	2	\,	2	)$	&	1	&	$(	2	\,	3	)$	&	1	\\
4	&	$(	0	\,	2	)$	&	1	&	$(	2	\,	2	)$	&	2	&	6	&	$(	6	\,	0	)$	&	1	&	$(	2	\,	0	)$	&	1	\\
4	&	$(	0	\,	2	)$	&	1	&	$(	1	\,	4	)$	&	1	&	6	&	$(	2	\,	2	)$	&	1	&	$(	2	\,	0	)$	&	1	\\
4	&	$(	4	\,	0	)$	&	1	&	$(	1	\,	1	)$	&	1	&	6	&	$(	2	\,	2	)$	&	1	&	$(	1	\,	5	)$	&	1	\\
4	&	$(	0	\,	2	)$	&	1	&	$(	0	\,	6	)$	&	1	&	6	&	$(	2	\,	2	)$	&	1	&	$(	1	\,	2	)$	&	1	\\
6	&	$(	6	\,	0	)$	&	1	&	$(	6	\,	4	)$	&	1	&	6	&	$(	2	\,	2	)$	&	1	&	$(	0	\,	4	)$	&	1	\\
6	&	$(	6	\,	0	)$	&	1	&	$(	6	\,	4	)$	&	2	&	6	&	$(	0	\,	0	)$	&	1	&	$(	0	\,	4	)$	&	1	\\
\end{tabular}
\end{center}
\label{sp04}
\end{table}

We employ a symmetry-guided concept, which allows the NCSpM model space to be down-selected to physically relevant symplectic bandheads, starting from bandheads of highest quadrupole deformation and lowest intrinsic spin. This means that the model space typically starts with only a few symplectic irreps, `vertically' extended to high $N_{\rm max}$, and is then expanded -- until convergence of results is achieved -- by adding more symplectic irreps, which introduce additional configurations within each \ho~subspace, thereby leading to a larger `horizontal' mixing. As the selection of additional symplectic irreps is based on the deformation of their bandheads, it is useful to note that the intrinsic quadrupole deformation of a bandhead is informed by its  \SU{3} labels according to the established mapping \cite{RosensteelR77b,LeschberD87,CastanosDL88}. Specifically, within an \ho-subspace, the deformation parameter $\beta^2$ of a bandhead is proportional to the expectation value of the  second-order Casimir invariant of \SU{3}:
\begin{equation}
{2\over 3} (\lambda_\sigma ^2+\mu_\sigma^2+\lambda_\sigma\mu_\sigma+3\lambda_\sigma+3\mu_\sigma).
\label{C2su3}
\end{equation}
Clearly, large $\lambda_\sigma$ and $\mu_\sigma$ imply large quadrupole deformation (large $\beta$), and bandheads are included in a model space in order of decreasing $\beta$, that is, decreasing values of Eq.(\ref{C2su3}).

This concept is informed by an {\it ab initio} study \cite{DytrychSBDV_PRL07,DytrychSDBV08_review} which used  the symplectic symmetry in an analysis of wave functions of $^{12}$C and $^{16}$O calculated with bare nucleon-nucleon ($NN$) interactions. Specifically, the outcome of this earlier {\it ab initio} study, corroborating the findings of preceding algebraic approaches \cite{RosensteelR77,DraayerWR84,Rowe85}, has revealed that typically symplectic many-body basis states built on only one or two bandheads of highest deformation and low spin suffice to represent a large fraction -- typically in excess of 80\% or more -- of the physics as measured by projecting { \it ab initio} NCSM results onto a symmetry-adapted equivalent basis  \cite{DytrychSDBV08_review}. Such a symplectic pattern has been also observed in {\it ab initio} symmetry-adapted no-core shell-model (SA-NCSM) studies of $^{6}$Li,  $^{6}$He, $^{8}$Be, and  $^{12}$C \cite{DraayerDLL_Erice11,DytrychLMCDVL_PRL12,REVIEW2016,DytrychMLDVCLCS11}.

In particular, the present study focusses on various model spaces for $^{12}$C consisting of symplectic irreps with bandheads of large quadrupole deformation and low intrinsic spin (Table \ref{tab:bandheads}), which allows results to be examined for convergence. Following the symmetry-guided concept, we first consider a model space consisting of the most deformed spin-0 0\ho, 2\ho, and 4\ho~bandheads  together with their symplectic excitations up through $N_{\max}=20$ with total dimensionality of $4.5\times10^3$ ($\mathfrak{C}$-1 in Table \ref{tab:bandheads}).  Then, the model space is expanded ``horizontally" to  $\mathfrak{C}$-2 (with total dimensionality  of $6.6\times10^3$), as well as to $\mathfrak{C}$-3 and $\mathfrak{C}$-4, which include higher-lying bandheads and bandheads  of decreasing deformation. The $\mathfrak{C}$-4 selection, for example, includes the symplectic irreps 0\ho(0\,4), 4\ho (12\, 0), 2\ho (6\, 2) and 6\ho (10\, 2) that have been identified to have the lowest mean-field energy based on shape-consistent mean-field considerations \cite{Rowe13}. We also consider the model space that includes all \SpR{3} irreps with low spin, spin-0 and spin-1, 0\ho~bandheads  with symplectic excitations up to $N_{\max}=20$ ($\mathfrak{C}$-5 in Table \ref{tab:bandheads}). This space consists of  the complete 0\ho~model space for $^{12}$C, excluding only the spin-2 part of the $(2\,0)$ 0\ho~ configuration, which is expected to be influenced by a spin-2 interaction, such as a tensor force  \cite{OtsukaSFGA, MyoUHTI}. Given the spin-0 and spin-1 nature of the model interaction we use, inclusion of the tensor force is outside of the scope of the present model, but will be considered in future studies. Nonetheless, as discussed next, the model interaction has been shown to yield results for $A=8$ to $A=24$ in close agreement with {\it ab initio} studies and experiment \cite{LauneyDDTFLDMVB13,DreyfussLDDB13,TobinFLDDB14}, with model space selections as small as $\mathfrak{C}$-1 and $\mathfrak{C}$-2 found to be sufficient.

\begin{table}[th]
\centering
\begin{tabular}{c|c|c|c|c|c|c}
Model 						& \multicolumn{2}{c|}{$0\ho$}							&$2\ho$		&$4\ho$		&$6\ho$		&$8\ho$ 			\\\cline{2-7}
Space						& $S=0$ 		& $S=1$		& $S=0$ 		& $S=0$ 		& $S=0$ 		& $S=0$ 			\\\hline\hline
$\mathfrak{C}$-1							&~$(0\,4)$~	& 			&~$(6\,2)$~	&~$(12\,0)$~	& 			&				\\\hline
$\mathfrak{C}$-2							&~$(0\,4)$~	&~$(1\,2)$~	&~$(6\,2)$~	&~$(12\,0)$~	& 			&				\\\hline
$\mathfrak{C}$-3							&~$(0\,4)$~	&~$(1\,2)$~	&~$(6\,2)$~	&~$(12\,0)$~	&~$(14\,0)$~	&~$(16\,0)$~		\\\hline
\multirow{4}{*}{$\mathfrak{C}$-4}				&~$(0\,4)$~	&~$(1\,2)$~ 	&~$(6\,2)$~	&~$(12\,0)$~	&~$(14\,0)$~	&~$(16\,0)$~			\\
							&			&			&~$(2\,4)$~	&~$(8\,2)$~	&~$(10\,2)$~	&				\\
							&			& 			&			&~$(4\,4)$~	&	 		&				\\
							&			& 			&			&~$(0\,6)$~	& 			&				\\\hline
\multirow{2}{*}{$\mathfrak{C}$-5}				&~$(0\,4)$~	&~$(1\,2)$~	&~$(6\,2)$~	&~$(12\,0)$~	&			&				\\
							&~$(2\,0)$~	&~$(0\,1)$~	&			&			&			&				\\\end{tabular}
\caption{\label{tab:bandheads} \SpR{3} irreps (specified by their bandhead labels) included in each of the model spaces considered. Each of model spaces $\mathfrak{C}$-1 through $\mathfrak{C}$-4 includes its preceding model space, while model space $\mathfrak{C}$-5 expands $\mathfrak{C}$-2 by including all spin-0 and spin-1 0\ho~ bandheads, which are, in fact, all the \SU{3} configurations that exist in the 0\ho~subspace. All model spaces extend up to $N_{\max}=20$.}
\end{table}

\subsection{Schematic many-nucleon interaction\label{Interaction}}
As discussed in Ref. \cite{DreyfussLDDB13}, we employ a microscopic many-body interaction, which enables large $N_{\max}$ no-core shell-model applications. This interaction utilizes two pivotal components:  a single-particle piece, consisting of the harmonic oscillator potential and a spin-orbit term, together with a collective piece, which enters through the quadrupole-quadrupole interaction, tied to a long-range expansion of the nucleon-nucleon central force $V(|\textbf{r}_i-\textbf{r}_j|)$ \cite{Harvey68}, 

\begin{eqnarray}
\label{effH}
H_{\gamma} &=& \sum_{i=1}^{A} \left(\frac{\textbf{p}_{i}^{2}}{2m}+\frac{m \Omega ^{2}\textbf{r}_{i}^{2}}{2} -\kappa l_{i}\cdot s_{i} \right) \nonumber \\
&& +\frac{\chi}{2} \frac{(e^{-\gamma (Q\cdot Q-\langle Q\cdot Q\rangle_{N_n})}-1)}{\gamma},
\end{eqnarray}
where \ho, $\kappa$, and $\chi$ are parameters, for which we use empirical estimates, and $\gamma \ge 0$ is the only adjustable parameter in the model (as discussed below). $H_{\gamma}$ and the mass quadrupole moment $Q_{(2M)}=\sum_{i=1}^{A}q_{(2M)i}=\sum_{i}\sqrt{16\pi /5}r_{i}^{2}Y_{2M}(\hat{\textbf{r}}_i)$ are given in terms of particle momentum and position coordinates relative to the center of mass. The quadrupole-quadrupole interaction, $\frac{1}{2}Q\cdot Q=\frac{1}{2}\sum_{i}q_i\cdot(\sum_{j}q_j)$ realizes the important interaction of each particle with the total quadrupole moment of the system. 
The average contribution $\langle Q\cdot Q\rangle_{N_n}$ of $Q\cdot Q$ for a given number of $N_n$ HO excitations \cite{RosensteelD85} introduces a considerable renormalization of the HO shell structure, and hence is removed in multi-shell studies \cite{CastanosD89}.

We use $\chi=\ho/(4\sqrt{ N_{\omega_f}N_{\omega_i} } )$ for a $\ME{f}{H_{\gamma}}{i}$ matrix element for a final (initial) many-body state, $f$ ($i$).
The decrease  of $\chi$ with $N_{\omega}$, to leading order in $\lambda_{\omega}/N_{\omega}$, has been shown by Rowe \cite{Rowe67} based on self-consistent arguments and used in an \SpR{3}-based study of cluster-like states of $^{16}$O \cite{RoweTW06}.  We also use the empirical estimates $\ho \approx 41/A^{1/3}$ and $\kappa \approx 20/A^{2/3}$ (e.g., see \cite{BohrMottelson69}).

The only adjustable parameter of the model is $\gamma$, which controls the presence of the many-body interactions in the model. The effective interaction (\ref{effH}) introduces hierarchical many-body interactions in a prescribed way. This ties directly to the interaction used in Ref. \cite{PetersonH80}, which is given as a polynomial in $Q$, and applied to the $^{24}$Mg ground-state ({\it gs}) rotational band. Indeed, higher-order terms in $Q\cdot Q$ of Eq. (\ref{effH}) become quickly negligible for a reasonably small $\gamma$. For example, we find that for $^{12}$C, besides $Q\cdot Q$, only one additional term -- $(Q\cdot Q)^2$ -- is sufficient for the ground-state band, with higher-order terms of the expansion being negligible. However, we find that the inclusion of terms up through $(Q\cdot Q)^4$ (third order in $\gamma$) is necessary for the Hoyle-state band \cite{DreyfussLDDB13}.

\subsection{Comparison to \emph{ab initio} no-core shell model}
\begin{figure}[th]
(a) $0^+_{gs}$ state in $^{12}$C 
\includegraphics[width=\columnwidth]{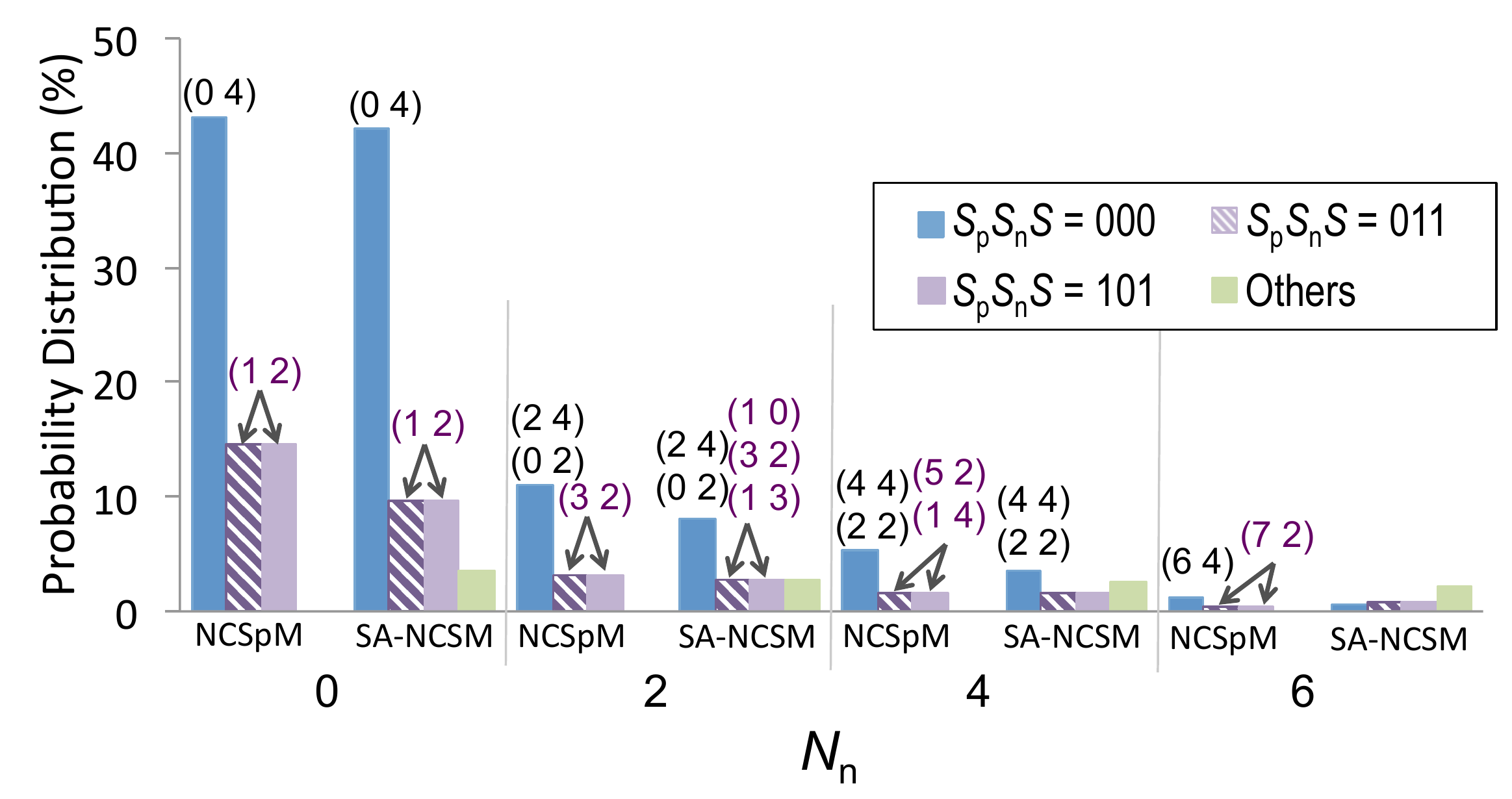}
(b) $4^+_{1}$ state in $^{12}$C 
\includegraphics[width=\columnwidth]{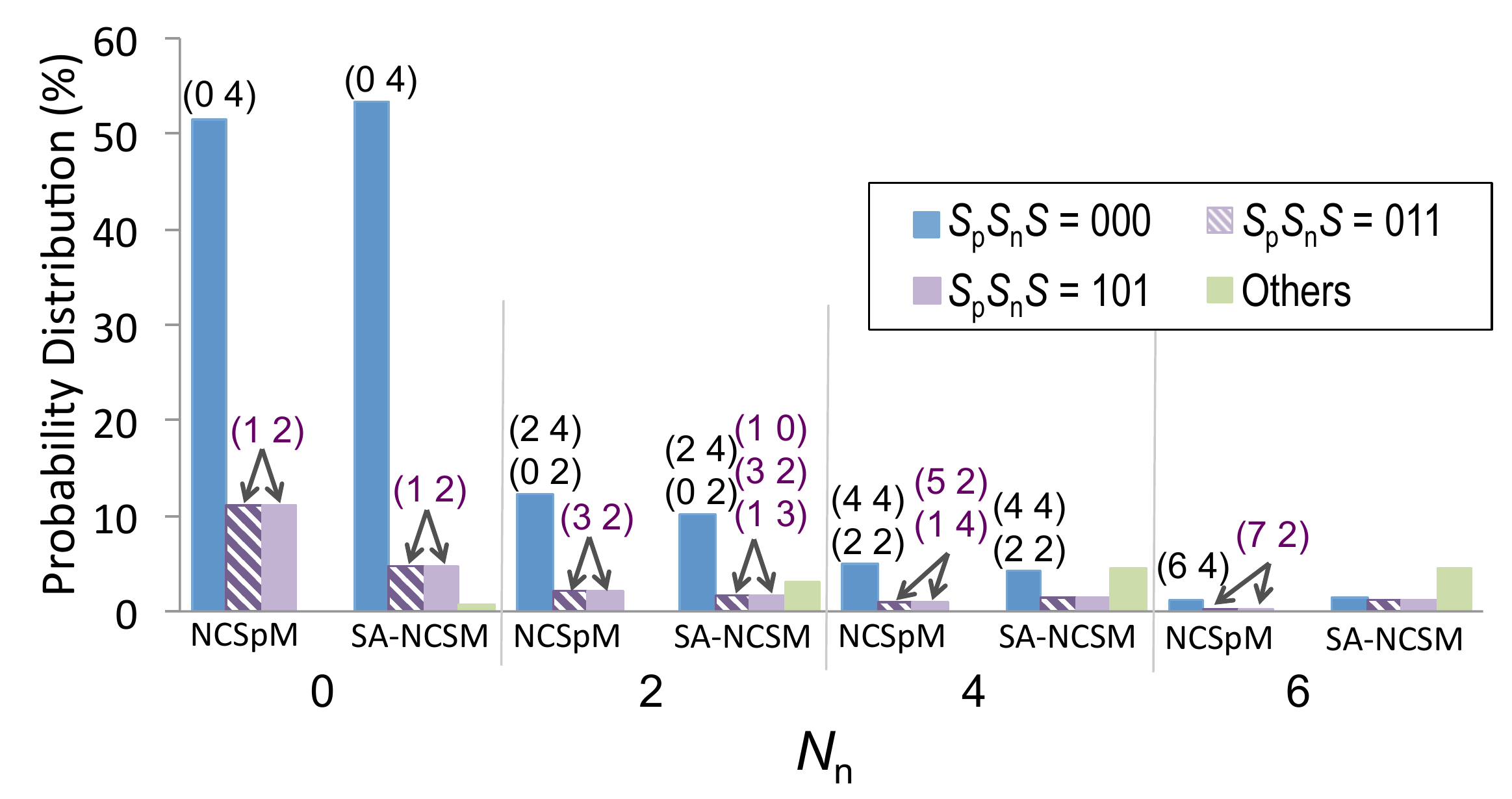}
\caption{ (Color online) Probability distribution for $^{12}$C across the $N_n$ total excitations of (a) the lowest $0^+$ state and (b) the lowest $4^+$ state as calculated by the NCSpM with $H_\gamma$ (left) in model space $\mathfrak{C}$-2 (see Table \ref{tab:bandheads}) and the {\it ab initio} SA-NCSM with the bare JISP16 $NN$ interaction (right). Both models are limited to an $N_{\max}=6$ model space for comparison. The dominant ``shapes", specified by $(\lambda\,\mu)$, are shown. This comparison is illustrated for the $0^+_{gs}$ and $4^+_{1}$ states, and the close similarity persists for the lowest $2^+$ state.}
\label{12C_Nmax6cmp}
\end{figure}
A comparison of our current results for $^{12}$C to {\it ab initio} outcomes is possible in smaller model spaces, for which {\it ab initio} NCSM calculations are feasible. For example, for the $gs$ rotational band, the $N_{\max}=6$ space appears to be reasonable  for both models (Fig. \ref{12C_Nmax6cmp}). In particular, we compare to wave functions obtained in the SA-NCSM \cite{DytrychLMCDVL_PRL12} with the bare JISP16 realistic interaction \cite{ShirokovMZVW07}. The SA-NCSM utilizes an \SU{3}-coupled basis, which yields eigenfunctions equivalent to the conventional NCSM wave functions \cite{NavratilVB00}, but realized in terms of the $(\lambda\,\mu)$ deformation labels, and hence, the deformed configurations that dominate the $^{12}$C wave functions can be straightforwardly studied.
\begin{figure*}[t]
\begin{center}
\includegraphics[width=0.61\textwidth]{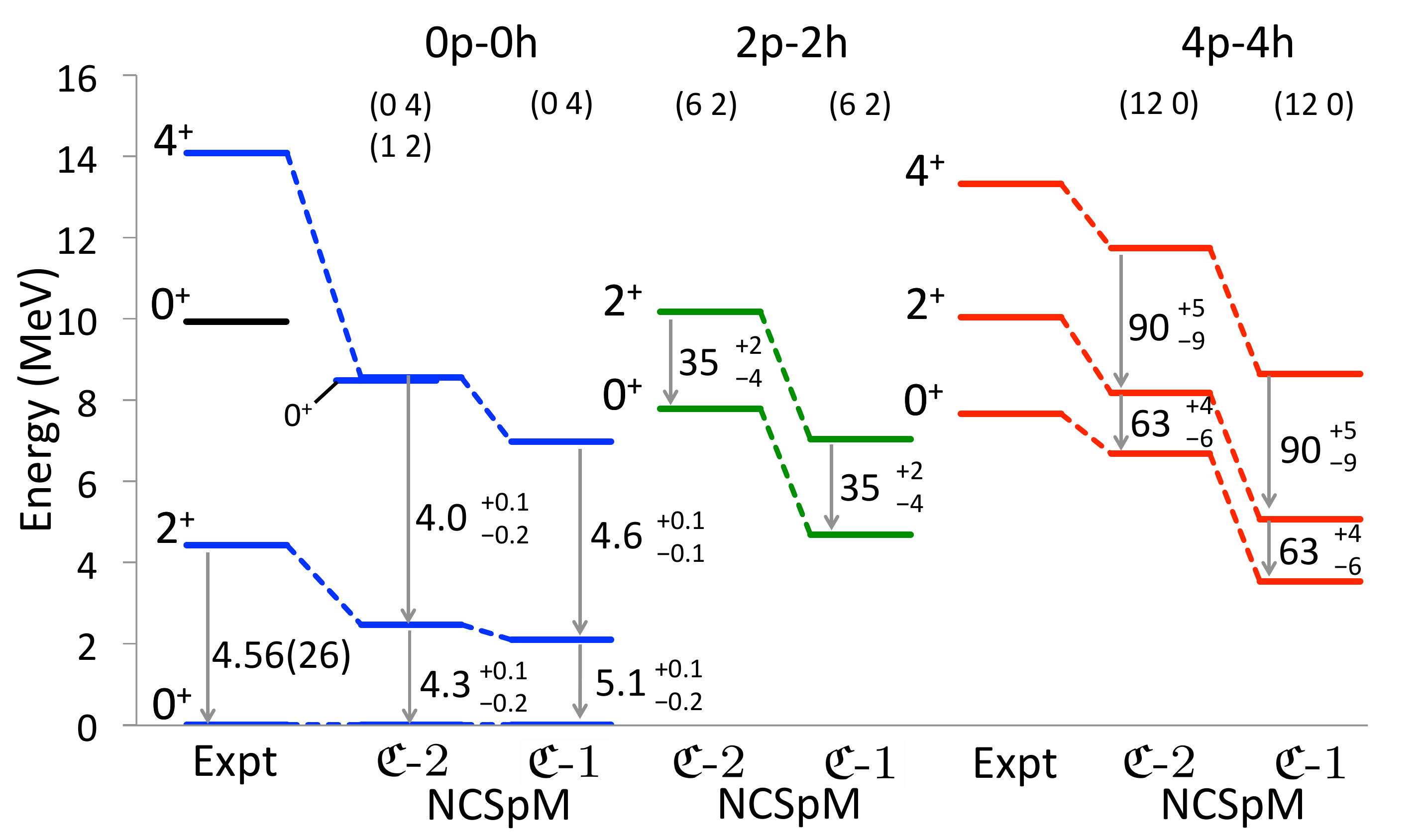}\\
\vspace{-1.3in}\hspace{1.8in}(Hoyle)\vspace{1.3in}
\caption{\label{fig:spectrum} (Color online) Energy spectrum for $^{12}$C calculated using the  NCSpM with symplectic irreps  starting at \ph{0} (blue, left), \ph{2} (green, center), and \ph{4} (red, right) bandheads and extending to $N_{\rm max}=20$, for model spaces $\mathfrak{C}$-1 and  $\mathfrak{C}$-2. Experimental data is from \cite{ASelove90}, except the latest results for $0_3^+$ \cite{Itoh11} and the states above the Hoyle state, $2^+$  \cite{Zimmerman13}  and $4^+$ \cite{Freer11}. The $B(E2;J\rightarrow J-2)$~transition rates are in W.u. with theoretical uncertainties estimated for a $\pm 60\%$ deviation of the Hoyle-state energy. Spectra calculated in model spaces $\mathfrak{C}$-3 and $\mathfrak{C}$-4 are the same as those shown for model space $\mathfrak{C}$-2.}
\end{center}
\end{figure*}

Consistent with the outcome of Refs.~\cite{DytrychLMCDVL_PRL12}  and \cite{LauneyDDTFLDMVB13b} (see, e.g., Fig.~1 in Ref. \cite{DytrychLMCDVL_PRL12} for $^6$Li and $^8$Be wave functions in $N_{\max}=8-10$), the {\it ab initio} $N_{\max}=6$ SA-NCSM results with the bare JISP16 realistic interaction for the $0^+$ {\it gs}, first $2^+$ and first $4^+$ states of  $^{12}$C reveal the dominance of the $0\ho$ component with the foremost contribution coming from the leading $(0\,4)$ $S=0$ irrep (see Fig.~\ref{12C_Nmax6cmp} for the {\it gs} and the $4^+_1$ state). Important \SU{3} configurations are then organized into structures with \SpR{3} symplectic symmetry, that is, the $(0\,4)$ symplectic irrep gives rise to dominant $(0\,2)$ and $(2\,4)$ configurations in the $2\hbar\Omega$ subspace and so on (see Fig.~\ref{12C_Nmax6cmp} and Table \ref{sp04}), and those configurations indeed realize the major components of each of the wave functions in this subspace. The next most important configuration is spin-1 $(1\,2)$ at 0\ho~ with its associated symplectic excitations (Fig.~\ref{12C_Nmax6cmp}). Among all possible configurations present in the SA-NCSM (total of $1.26\times 10^6$ for $J=0$ in $N_{\max}=6$), only the states of the $(0\, 4)$ and then  $(1\, 2)$ symplectic cones appear dominant. This further confirms the significance of the symplectic symmetry to nuclear dynamics. The outcome points to the fact that the relevant model space can be systematically determined by down-selecting to important spin configurations in lower subspaces while expanded to include a manageable  set of symplectic configurations in the higher $N_{\max}$ regime.

Furthermore, we find a close similarity between complete {\it ab initio} SA-NCSM results and the NCSpM wave functions of the $^{12}$C  $gs$ rotational band, calculated with $H_{\gamma}$ of Eq. (\ref{effH}) for $\gamma=1.71\times 10^{-4}$ and symplectic irreps of model space $\mathfrak{C}$-2 (Fig.~\ref{12C_Nmax6cmp}). NCSpM and SA-NCSM calculations are performed for  $\ho=18$  and an $N_{\max}=6$ model space. The two models show close agreement of the probability distribution, including the \SU{3} content of the wave functions. This suggests that the interaction used in the NCSpM has effectively captured a major portion of the underlying physics of the realistic interaction important to the low-lying nuclear states.

\section{Results and discussions}

The NSCpM utilizes Bahri's symplectic computational code \cite{bahri94rme}, which uses Draayer \& Akiyama's \SU{3} package \cite{DraayerSU3_1}. The symmetry-mixing spin-orbit term is calculated in the SA-NCSM \cite{DytrychLMCDVL_PRL12}. This term is applied only to the symplectic bandheads and provides a `horizontal' mixing of the symplectic irreps.

The model successfully reproduces the ground-state and Hoyle-state rotational bands in $^{12}$C \cite{DreyfussLDDB13}, where both rotational features and $\alpha$-cluster substructures are shown to emerge in the fully microscopic $N_{\rm max}=20$ no-core shell-model framework, as suggested by the reasonably close agreement of  the model outcome with experiment and {\it ab initio} results in smaller spaces. While the model includes an adjustable parameter, $\gamma$, this parameter only controls the presence of many-nucleon interactions, and hence, introduces an additional, but very limited, degree of freedom. The entire many-body apparatus is fully microscopic and no adjustments are possible. We find that, as $\gamma$ varies, there is only a small window of possible $\gamma$ values around $\gamma =1.71\times 10^{-4}$ which, for large enough $N_{\max}$, closely reproduce the relative positions of the three lowest $0^+$ states in $^{12}$C and associated measured observables, discussed below.  The model has been also applied to low-lying states of other nuclei, such as $^{8}$Be and $sd$-shell nuclei \cite{LauneyDDTFLDMVB13,TobinFLDDB14}, without any further  parameter adjustment. In particular, using the same $\gamma=1.71\times 10^{-4}$ as determined for $^{12}$C,  we have described selected low-lying states in  $^{8}$Be in an  $N_{\rm max}=24$  model space with only 3 spin-0 $0\ho$ $(4\, 0)$, $2\ho$  $(6\, 0)$, and  $4\ho$ $(8\, 0)$ symplectic irreps \cite{LauneyDDTFLDMVB13}, as well as the ground-state rotational band of heavier nuclei, such as $^{20}$O, $^{20}$Ne, $^{22,24}$Ne, $^{20,22,24}$Mg, and $^{24}$Si, using $N_{\rm max}=12$  model spaces \cite{TobinFLDDB14}.   

\subsection{Clustering and collectivity in $^{12}$C \label{Hoyle}}
In this section, we focus on the ground state and Hoyle state in $^{12}$C, along with their rotational bands, and study the dependence of the NCSpM results on the model space and the model parameters $\gamma$. As described above, we use $H_{\gamma}$ with $\gamma = 1.7\times 10^{-4}$ along with, for $A=12$, $\hbar \Omega = 18$ MeV and $\kappa = 3.8$ MeV.

Analysis of the results shows that model space $\mathfrak{C}$-1, consisting of irreps built upon the spin-0 $0\ho$ \ph{0} $(0\, 4)$, the $2\ho$ \ph{2} $(6\, 2)$, and the $4\ho$ \ph{4} $(12\, 0)$ bandheads, is capable of bringing the Hoyle state down in energy (Fig.~\ref{fig:spectrum}, last column). For this model space, we observe three low-lying $0^+$ states below 10 MeV, and their rotational bands (e.g., $0^+$, $2^+$, and $4^+$): the \ph{0} ground state (Fig. \ref{fig:spectrum}, first column), the \ph{4} $0^+$ state that tracks with the Hoyle state, and a \ph{2} (Fig. \ref{fig:spectrum}, middle column) above the \ph{4} $0^+$ state. However, this model space yields a compressed energy spectrum. We note that the spin-orbit interaction, being a tensor operator of spin 1, does not mix spin-0 irreps. Hence, for this model space, the spin-orbit term has no effect (equivalent to $H_\gamma$ with $\kappa=0$) and the $H_\gamma$ eigenstates consist of a single symplectic irrep.

With the expansion of the model space by only one spin-1 irrep (model space $\mathfrak{C}$-2), the $N_{\max}=20$ NCSpM energy spectrum is improved and found to lie reasonably close to the experimental data (Fig.~\ref{fig:spectrum}, see $\mathfrak{C}$-2) \cite{DreyfussLDDB13}. The \SpR{3}-nonpreserving spin-orbit term mixes the spin-0 $(0\, 4)$ and spin-1 $(1\, 2)$ irreps for all $J^{\pi}=0^{+}$, $2^{+}_1$, and $4^{+}_1$, which results in a more realistic energy spacing between the excited states. Specifically, we see the $gs$ separating from the higher-lying $0^{+}$ states, and a slight stretching in the $gs$ rotational band. This agrees with early cluster models that showed similarly compressed spectra, which were corrected through allowing for $\alpha$-cluster dissociation due to a spin-orbit force, as discussed in Ref.\cite{Kanada98}. The inclusion of this additional irrep introduces another low-lying \ph{0} $0^+$ state (Fig. \ref{fig:spectrum}, first column), which -- along with the \ph{2} $0^+$ state -- lies close to the broad 10-MeV $0^+$ resonance observed in $^{12}$C .

In the present model, the spin-orbit interaction is turned on only among the bandheads of the symplectic irreps, up to $N_{\rm max}=4$ for the $\mathfrak{C}$-2 model space (and $N_{\rm max}=8$ for $\mathfrak{C}$-4), which results in the mixing of basis states within $S=0$ and $S=1$ irreps up to  $N_{\rm max}=20$ (see the NCSpM results shown in Fig. \ref{12C_Nmax6cmp} for $N_{\rm max}=6$ and the $^{12}$C ground state). These calculations are performed in the SA-NCSM, referenced above, which is ideal for the symplectic bandheads under consideration, because they are equal to the corresponding \SU{3} basis states of the SA-NCSM. The full accounting of the spin-orbit interaction is estimated, at the most, to render additional mixing of about 0.2\% for $(6\, 2)$,  $4\times 10^{-4}$\% for $(12\, 0)$, and $11\%$  for $(1\, 2)$ and $(0\, 4)$ to the $^{12}$C $gs$, while increasing the corresponding $0^+$ state energies by only a few MeV without affecting their order. That the bandheads provide a reasonable account of the spin-orbit effect stems from an important feature of  the $l\cdot s$ operator -- it is a spin-1 0\ho $(1\,1)$ \SU{3} tensor and only mixes certain configurations within the irreps. Specifically, the main contribution to the spin-orbit matrix elements  between the $(1\, 2)$ irrep and the $(6\, 2)$ irrep, or the $(12\, 0)$ irrep,  comes from higher-$N_n$ configurations where the $(1\, 2)$ probability amplitudes are already small, 1-8\% (see Fig. \ref{12C_Nmax6cmp}). In addition, mixing to the $(6\, 2)$ and $(12\, 0)$ irreps is not allowed by \SU{3} selection rules for the most dominant configurations in these irreps, and it involves only configurations of probability amplitudes of less than 0.5\% for $(6\, 2)$ and 0.02\% for $(12\, 0)$. This results in negligible effects on the states and associated energies. In addition, the bandheads of the $(0\,4)$ and $(1\,2)$ irreps constitute a major component of the wave functions, which is $\sim 70\%$ of the $0^{+}_{gs}$, $2^{+}_{1}$, and $4^{+}_{1}$ states.
\begin{table}[t]
\caption{Transition rates, $B(E2)$ in W.u.~and $M(E0)$ in $e\,$fm$^2$, as well as rms matter radii ($r_{rms}$) in fm and the electric quadrupole moment in $e\,$fm$^2$ obtained by the NCSpM with $H_{\gamma}$ in model spaces $\mathfrak{C}$-1 and  $\mathfrak{C}$-2 (with $\mathfrak{C}$-2 results coinciding with those for model spaces $\mathfrak{C}$-3 \& $\mathfrak{C}$-4), as well as for a $1.7\%$ mixing of the $(12\,0)$ irrep into the $(0\,4)$ irrep (see text for details). Experimental values are shown in the rightmost column.}
\begin{center}
{\footnotesize
\begin{tabular}{lccccc}
\hline\hline
										& \multicolumn{2}{c}{$\mathfrak{C}$-1} 	 	&  \multicolumn{2}{c}{$\mathfrak{C}$-2}   		&  Expt.		\\ 
										& NCSpM	& Mixing		&   NCSpM	&  Mixing				&  Ref.~\cite{ASelove90}				\\\hline
$B(E2;\,2^+_1\!\rightarrow\!0^+_{gs})$  			& 5.12 		& 4.37	  		&  4.3			&  3.64				& 4.65(26) 			\\
$B(E2;\,0^+_2\!\rightarrow\!2^+_{1})$  			&  0      		& 8.7  			&  0				&  8.4				& 8.0(11)				\\
$B(E2;\,2^+_2\!\rightarrow\!0^+_{2})$  			& 63.2  		& 60.5	  		&  63.2				&  60.5				& N/A			\\
$M(E0;\,0^+_2\!\rightarrow\!0^+_{gs})$ 			& 0			& 2.04			&  0				&  2.1				& 5.4(2)				\\
$r_{rms}~0^+_{gs}$  						& 2.44 		& 2.45		  	&  2.43(1)			&  2.44				& 2.43(2)	 			\\
$r_{rms}~0^+_{2}$	(Hoyle)					& 2.93 		& 2.92		  	&  2.93(5)			&  2.92				& 2.89(4)	 			\\
$Q_{2^+_1}$								& 6.63 		& 6.17	  		&  5.9(1)			&  5.44				& +6(3)	 			\\
\hline\hline
\end{tabular}
}
\end{center}
\label{tab:BE2}
\end{table}%

Of particular note is the $2^+$ state, calculated by the NCSpM as a rotational excitation 1.51 MeV above the second $0^+$ state (see Fig.~\ref{fig:spectrum}, last column). Morinaga was the first to suggest that this $2^+$ state, which he estimated to be at 9.7 MeV, could be a member of a Hoyle-state rotational band \cite{PRMorinaga}. The existence of a $2^+$ state near this energy has important implications for astrophysical reaction rates \cite{Fynbo05}, and has been the subject of many experimental studies since Morinaga first suggested it as a mechanism to probe the structure of the Hoyle state. More recent experimental studies have given rise to much debate surrounding the $2^+_2$ state: inelastic $^{12}$C($\alpha, \alpha'$) and $^{12}{\rm C}({\rm p}, {\rm p}')$ scattering reactions showed evidence for an excited $2^+_2$ around 9.6-11 MeV \cite{ZimmermanDFGS11,Freer0709, Freer2010, Itoh11}, but studies of the $\beta$-decay of $^{12}$N and $^{12}$B found no evidence for the existence of a $2^+$ state below 10 MeV \cite{Hyldegaard10,NPatel13}. The NCSpM first identified a low-lying $2^+$ state as a part of the $0^+_2$ rotational band at 10.68 MeV \cite{LauneyDDTFLDMVB13b}, which used model space $\mathfrak{C}$-1 and a rescaling factor. A subsequent study used the $^{12}$C($\gamma$, $\alpha_0$)$^{8}$Be reaction, and identified the $2^+_2$ state at 10.03(11) MeV with a total width of 800(130) keV \cite{Zimmerman13}, or approximately 2.4 MeV above the Hoyle-state energy. 
For comparison, recent {\it ab initio} $N_{\max}=8$ NCSM calculations, while achieving a remarkable reproduction of the $gs$ rotational band, yield the second low-lying $0^+$ and $2^+$ states around 13 MeV and 15 MeV, respectively \cite{RothLCBN11}, which are thus believed to be associated with higher-lying states of spin-parity $0^+$ and $2^+$.
Here, we also identify a low-lying $4^+$ state at 11.7 MeV (see Fig. \ref{fig:spectrum}), which tracks with experimental identification of a low-lying $4^+$ state believed to be in the Hoyle-state rotational band \cite{Freer11}. 
\begin{figure*}[th!]
\centering
(a) \hspace{3in} (b)\\
\includegraphics[width=0.42\textwidth]{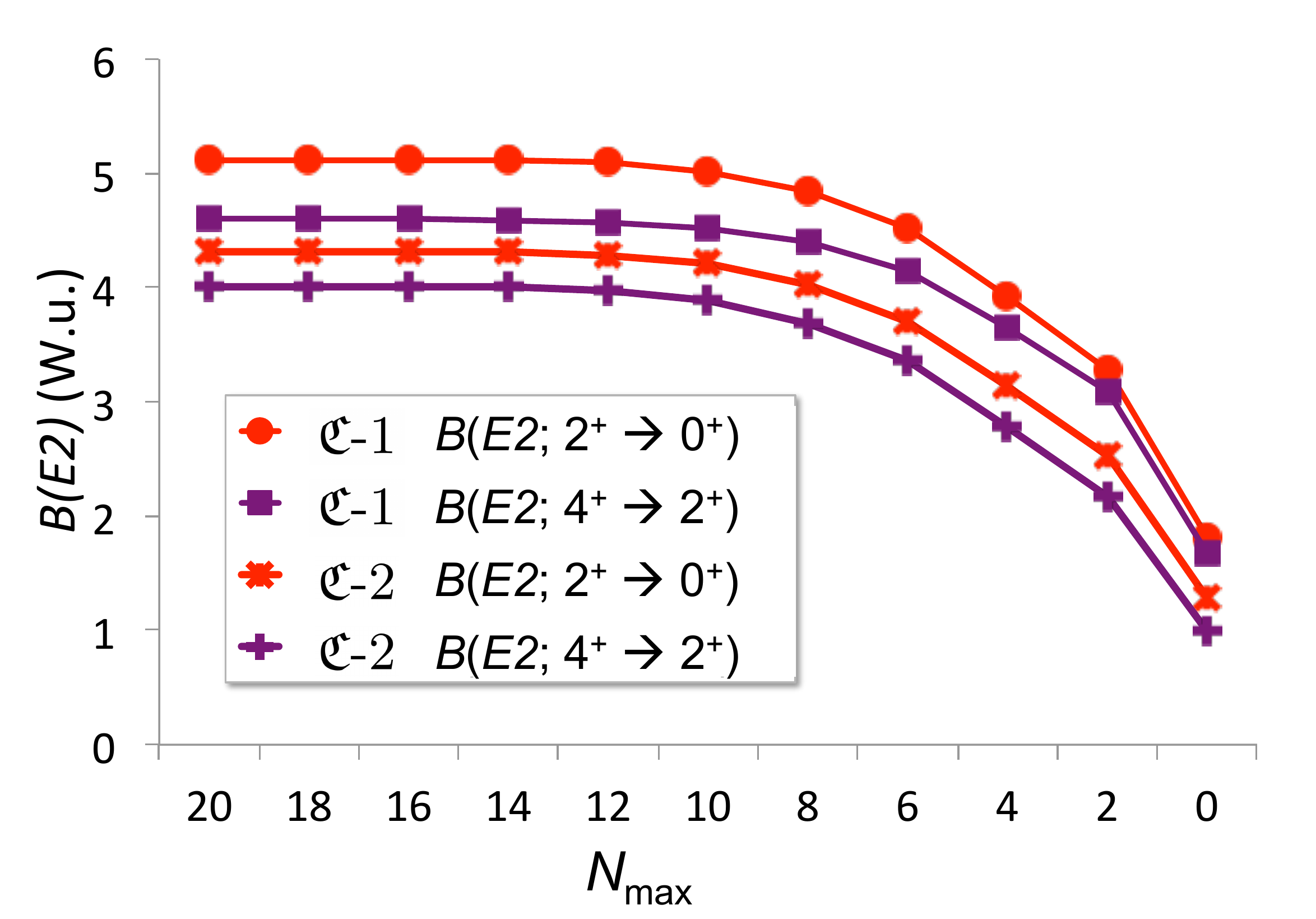}
\includegraphics[width=0.42\textwidth]{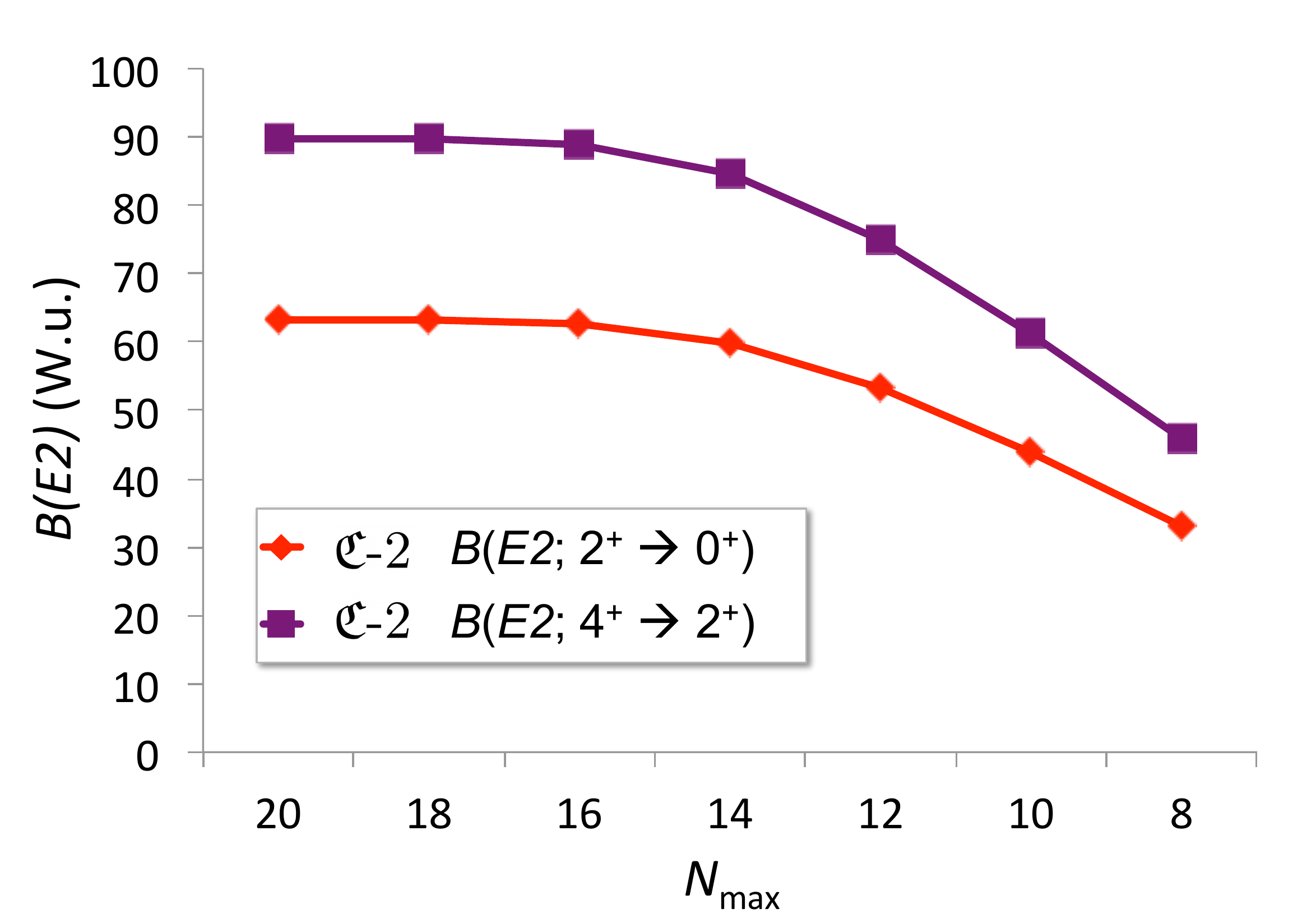}\\
(c) \hspace{3in} (d)\\
\includegraphics[width=0.42\textwidth]{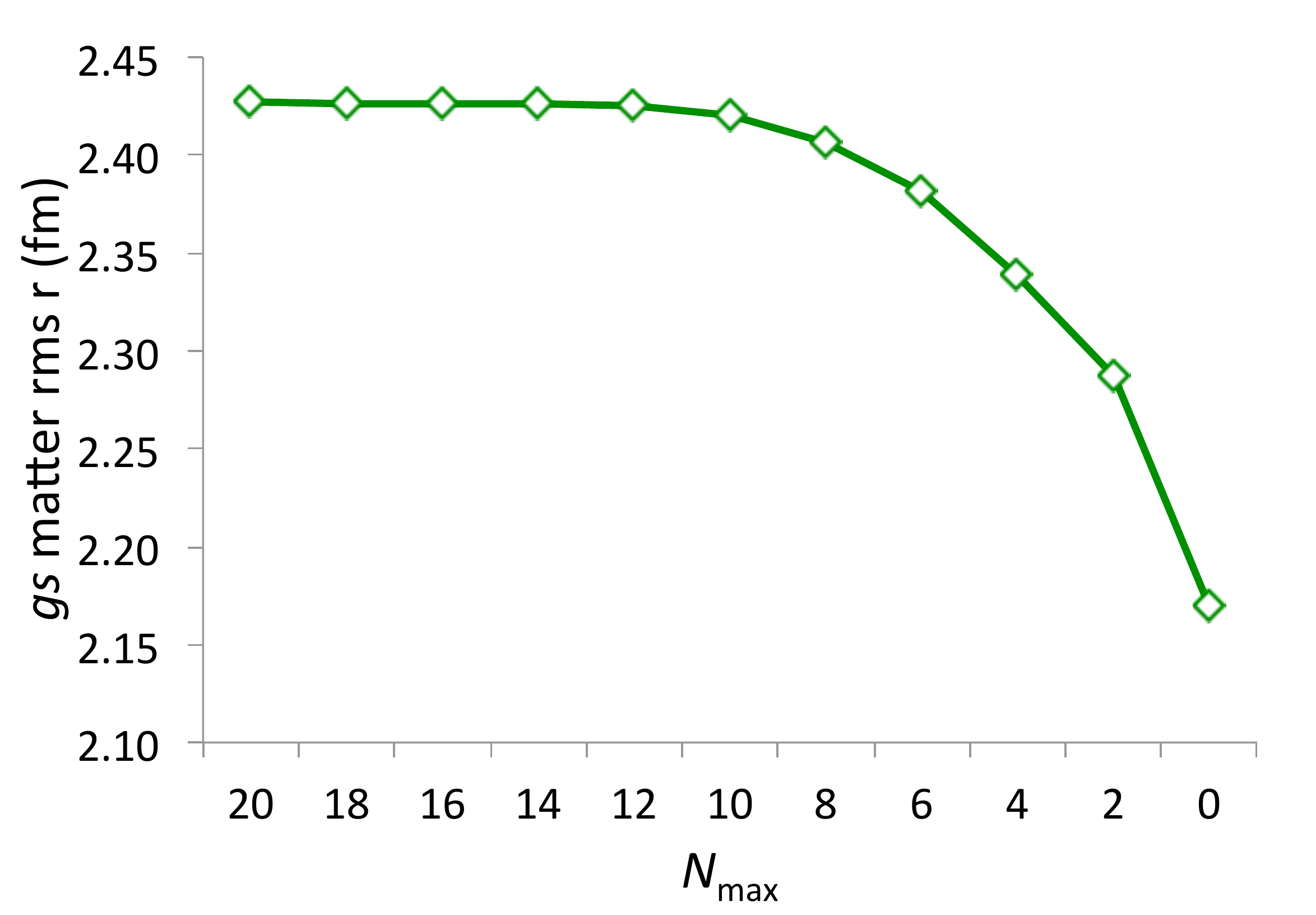}
\includegraphics[width=0.42\textwidth]{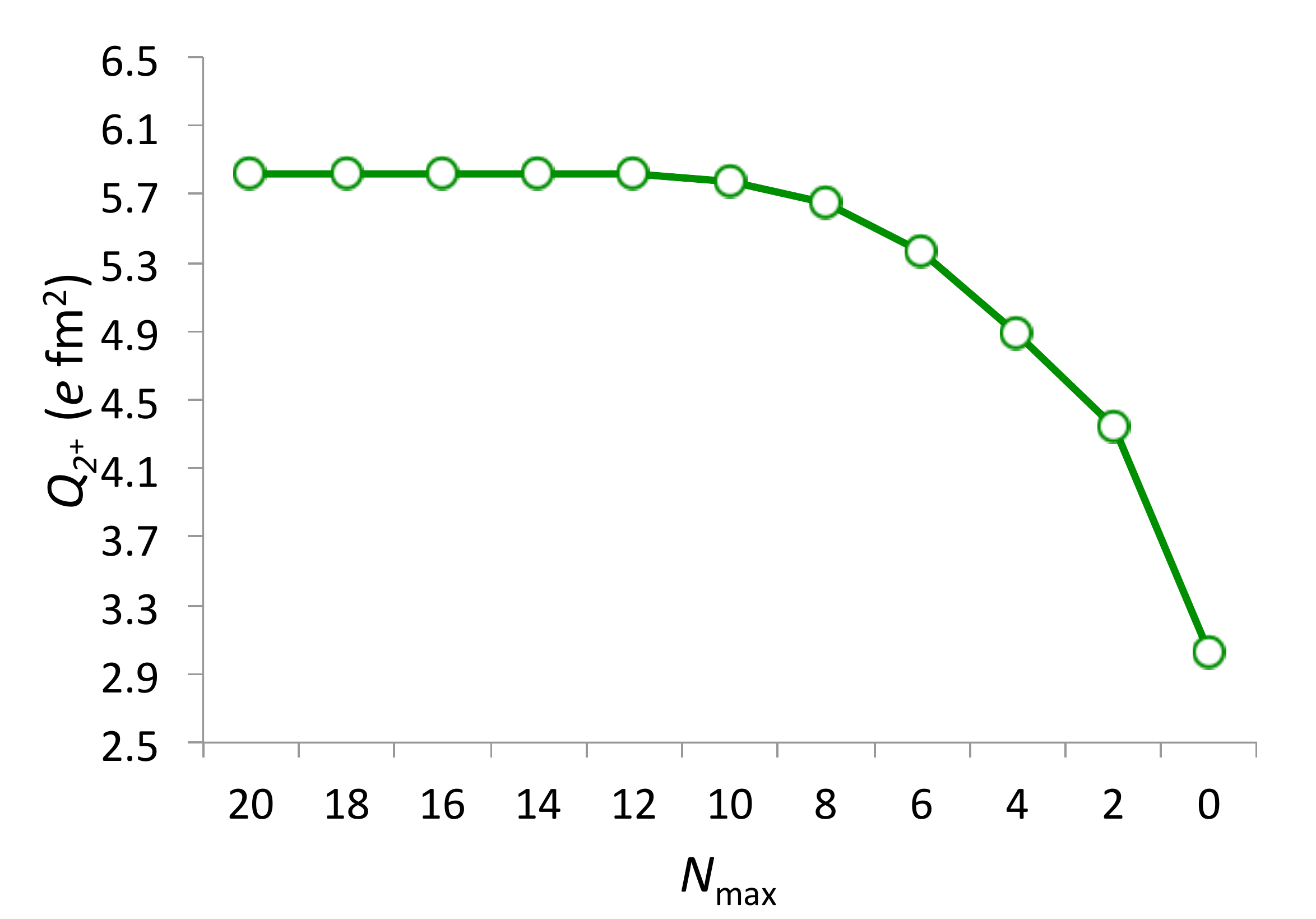}
  \caption{ (Color online) Dependence of NCSpM ($\gamma =1.71\times 10^{-4}$) on $N_{\max}$ for (a) $B(E2)$ of the {\it gs} rotational band in model spaces $\mathfrak{C}$-1 and $\mathfrak{C}$-2 (see Table~\ref{tab:bandheads}), as well as (b) $B(E2)$ of the Hoyle-state rotational band, (c) $gs$ point-particle matter rms radius, and (d) the electric quadrupole moment for $2_{1}^+$ in model space $\mathfrak{C}$-2 (with results for $\mathfrak{C}$-3 and $\mathfrak{C}$-4 identical to those of $\mathfrak{C}$-2).}
\label{NmaxConvergence}
\end{figure*}

The NCSpM is also used to study observables of $^{12}$C, such as $B(E2)$ transition strengths, $Q_{2^+_1}$, and matter rms radii for the $gs$ and Hoyle state. Comparison of results for model spaces $\mathfrak{C}$-1 and $\mathfrak{C}$-2  (see columns 2 and 4 in Table~\ref{tab:BE2}) shows slight differences, implying that the spin-orbit interaction  has only a small effect on these observables. While the in-band transition strengths are quite reasonable, a nonzero $B(E2;\,0^{+}_2\!\rightarrow\!2^{+}_1)$ value can only result from mixing of symplectic irreps, which requires an interaction with an \SpR{3} symmetry-breaking term beyond the spin-orbit interaction. To examine a possible mixing of the \ph{4} (12\, 0) irrep into the ground state, we consider an ad-hoc mixing of the \ph{0} and \ph{4} irreps, which is equally applied to all the states within each irrep. However, we find that an extremely small mixing, $1.7\%$, of the $(12\,0)$ irrep into the \ph{0} irreps of the $gs$ rotational band is sufficient to realize the observed $B(E2)$ rates and to yield results consistent with the $M(E0)$ experimental value (Table \ref{tab:BE2}). The results indicate that while the mixing has some effect on the collectivity within the $gs$ rotational band, the matter rms radii for the ground and Hoyle states remain unaffected.

\begin{table}[th]
\centering
\begin{tabular}{lcc}
&\multicolumn{2}{c}{Energy (MeV)}\\
$N_{\sigma}(\lambda_{\sigma}\,\mu_{\sigma})$&$\mathfrak{C}$-3 &$\mathfrak{C}$-4 \\
\hline
~$2\ho(2\,4)$				&&30.68	\\
~$4\ho(8\,2)$				&&31.94	\\
~$4\ho(4\,4)$				&&55.61	\\
~$4\ho(0\,6)$				&&70.53	\\
~$6\ho(14\,0)$				&34.21 &34.21	\\
~$6\ho(10\,2)$				&&57.56	\\
~$8\ho(16\,0)$				&63.12&63.12	\\\hline
\end{tabular}
\caption{\label{tab:lowzero} Low-lying $0^+$ states calculated as the lowest $0^+$ state for each \SpR{3} irrep specified by its bandhead in the table and for the model spaces $\mathfrak{C}$-3 and $\mathfrak{C}$-4 (see Table~\ref{tab:bandheads}). Energies are reported with respect to the ground state in MeV. For comparison, the Hoyle-state energy given by the lowest  $0^+$ state within the $4\ho(12\,0)$ irrep is 6.66 MeV.}
\end{table}
\begin{figure*}[th!]{}
\centering
(a) \hspace{3in} (b)\\
\includegraphics[width=0.49\textwidth]{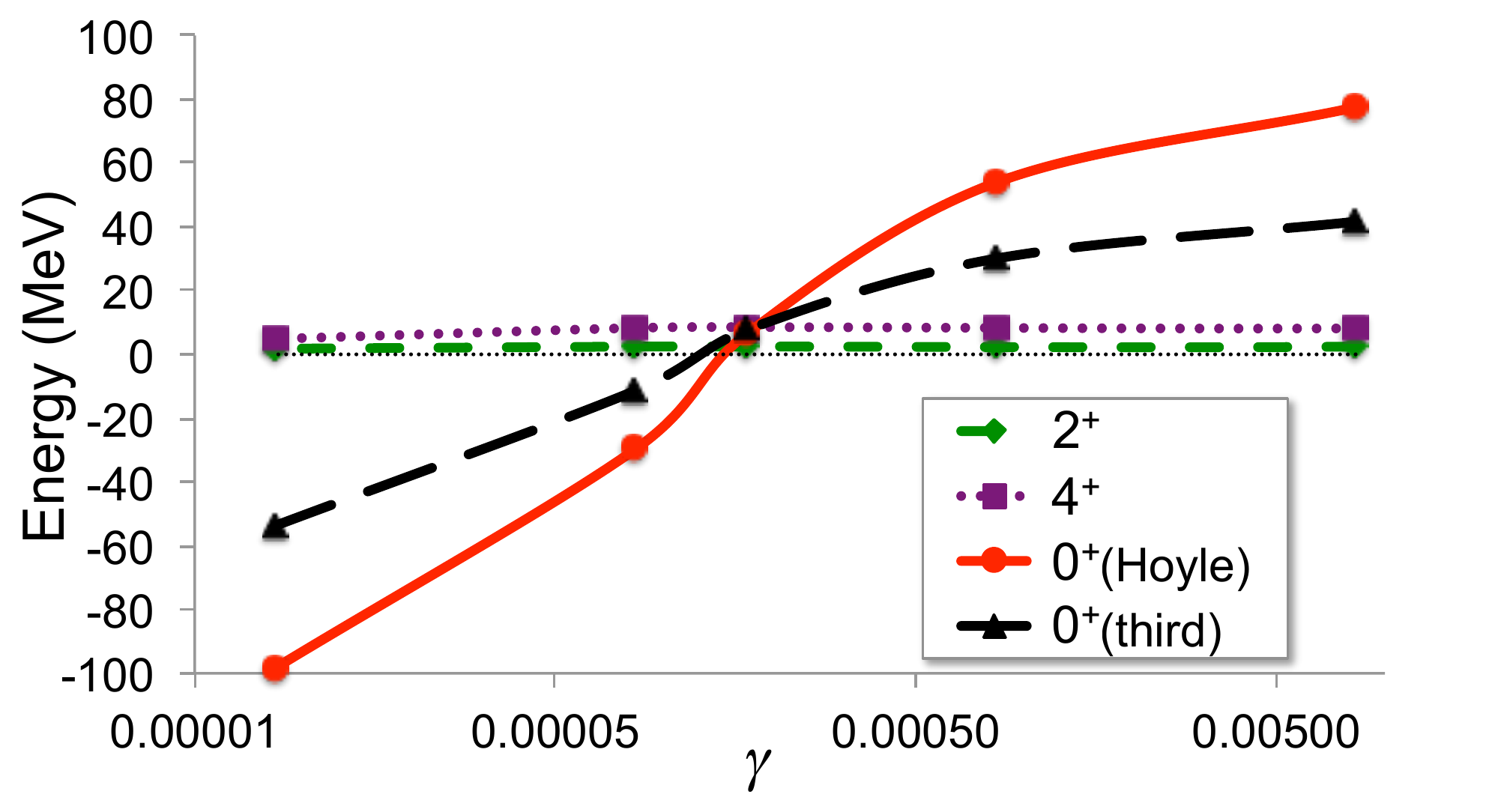} 
\includegraphics[width=0.49\textwidth]{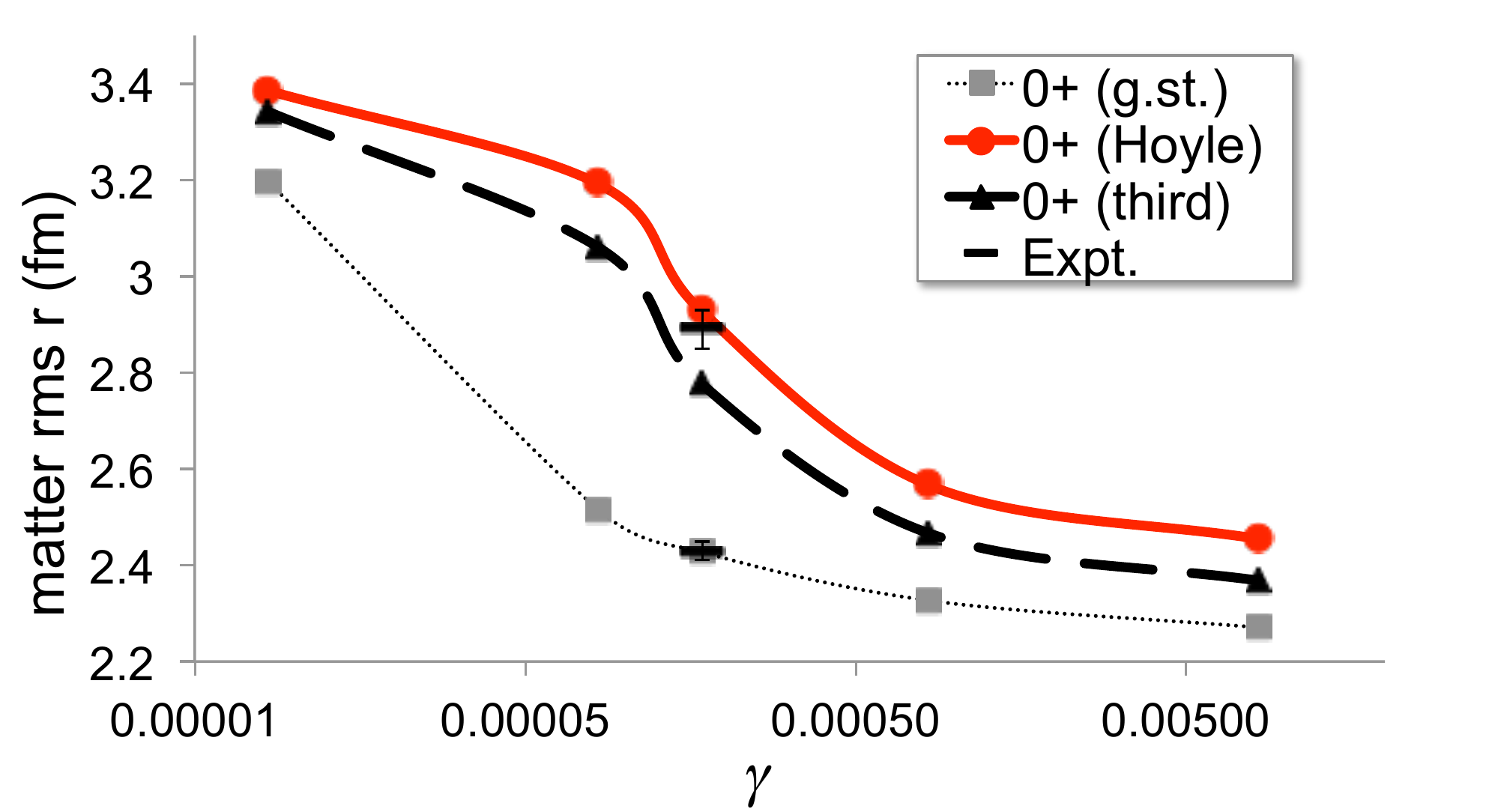}\\
(c) \hspace{3in} (d)\\
\includegraphics[width=0.49\textwidth]{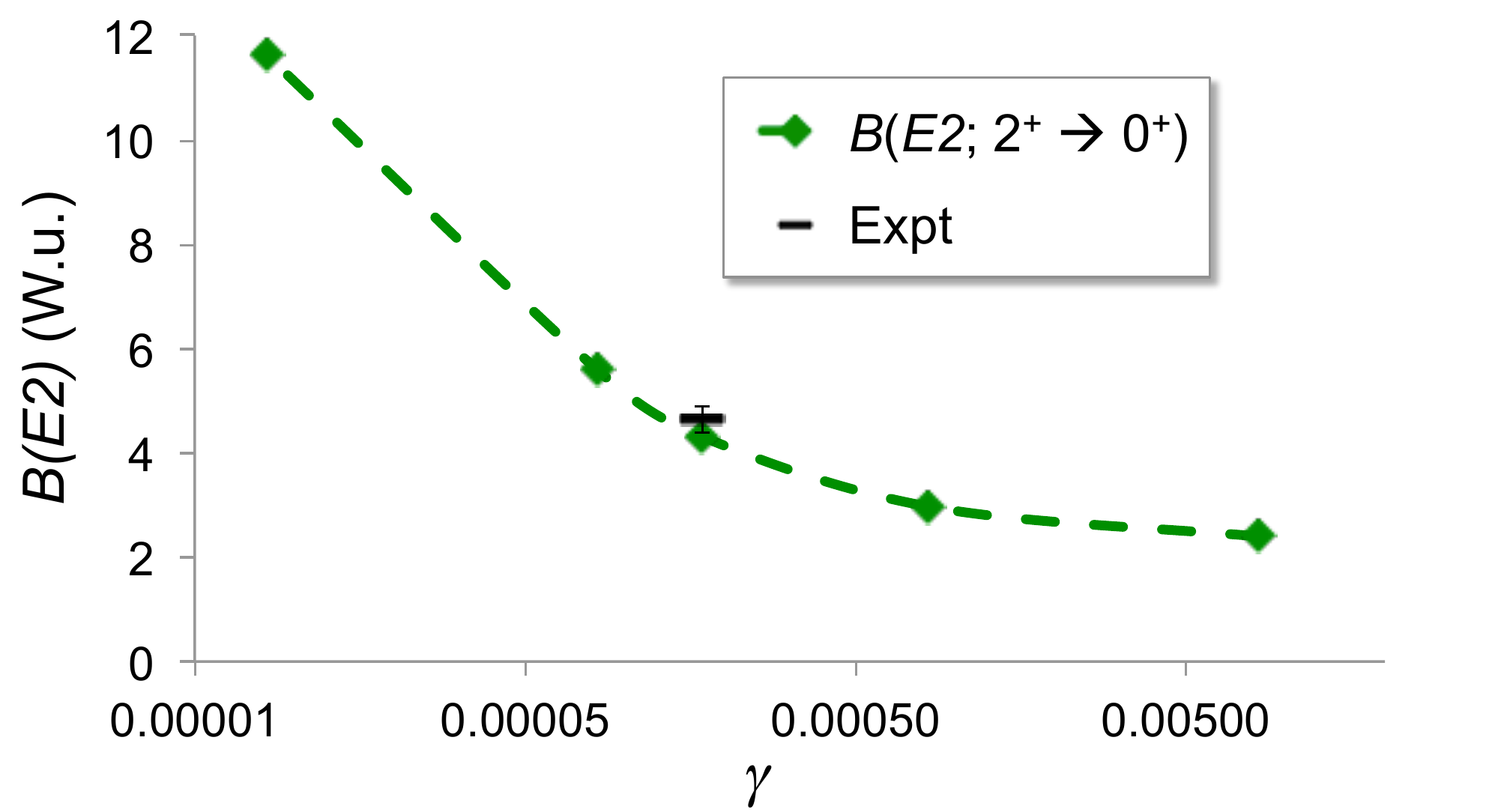}
\includegraphics[width=0.49\textwidth]{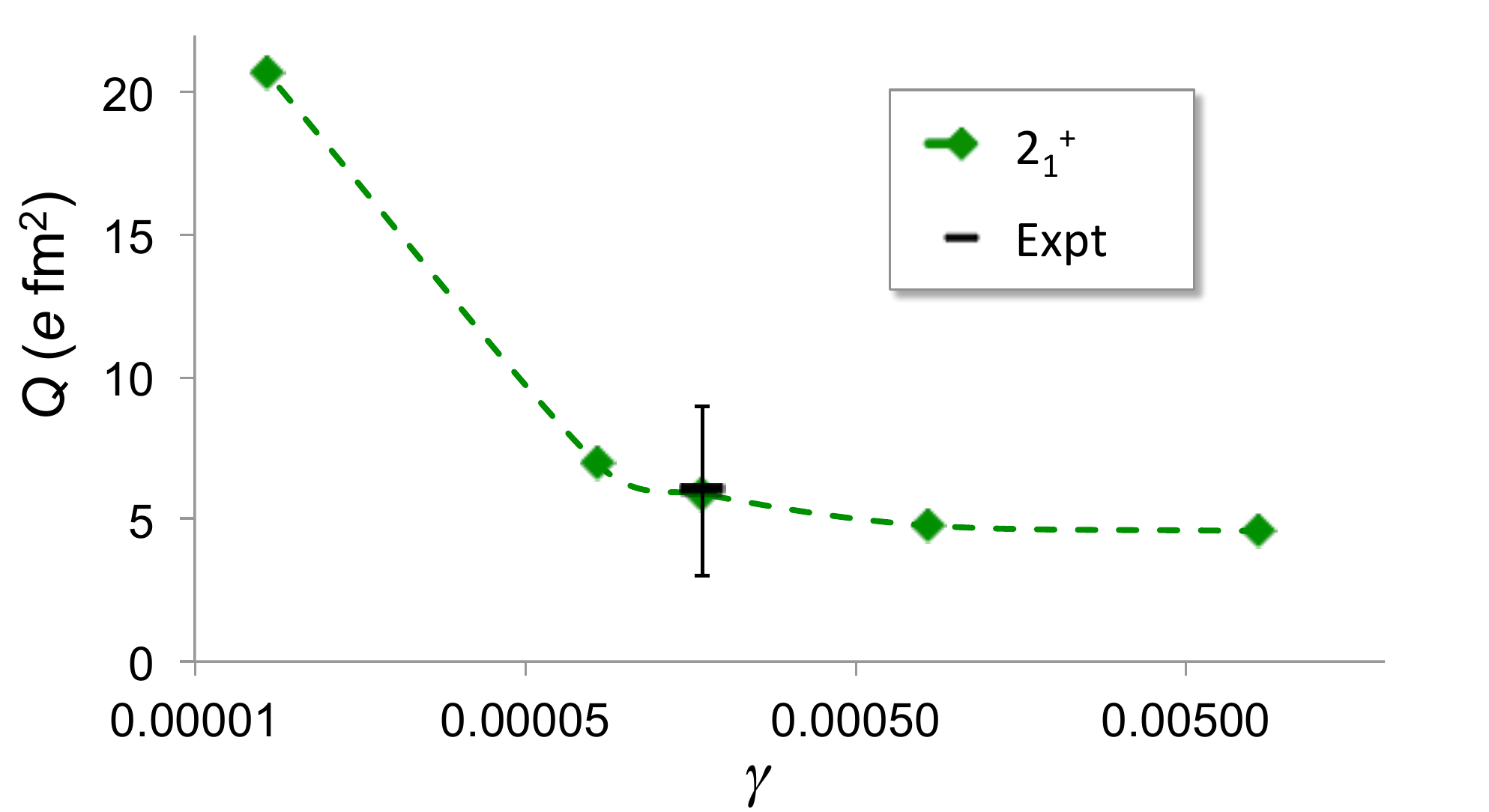}
  \caption{(Color online) Dependence of the $^{12}$C NCSpM energy spectrum on the $\gamma$ model parameter for $N_{\rm max}=20$ in model space $\mathfrak{C}$-2 (see Table~\ref{tab:bandheads}). Available experimental values are shown for (b), (c), and (d) and compared to the NCSpM results quoted in Table \ref{tab:BE2} for $\gamma=1.71\times 10^{-4}$.}
  \label{12C_gammaspanNmax20}
\end{figure*}

\noindent \textbf{Dependence on horizontal expansion} --
As shown above, reasonable results for $^{12}$C are obtained using the $\mathfrak{C}$-2 model space. We examine a possible dependence of the outcome as more symplectic irreps are added into the model space by considering $\mathfrak{C}$-3 and $\mathfrak{C}$-4 (Table~\ref{tab:bandheads}). This leads to more configurations within each ``horizontal" HO shell. We find that all the $\mathfrak{C}$-2 results presented in Fig. \ref{fig:spectrum} and Table \ref{tab:BE2} remain unaltered, and that no additional low-lying $0^+$ states are introduced to the $^{12}$C spectrum with the inclusion of the most deformed $S=0$ bandheads at $6\ho$ and $8\ho$ as in model space $\mathfrak{C}$-3, nor with the inclusion of $S=0$ bandheads of decreasing deformation, as in model space $\mathfrak{C}$-4 (Table \ref{tab:lowzero}). Thus, we find the results to be converged with respect to a horizontal expansion of the model space. Model space $\mathfrak{C}$-5 does produce an additional low-lying, $(2\, 0)$-dominated $0^+$ state below the Hoyle-state energy. A similar state appears at 15 MeV in {\it ab initio}  SA-NCSM calculations for the complete $N_{\rm max}=8$ model space. However, with a radius almost equal to that of the ground state ($2.41$ fm) and a very weak monopole transition strength ($0.29~e$fm$^2$), this is not a viable candidate for the Hoyle state.

\noindent \textbf{Dependence on vertical expansion} -- 
A study of the effect of the $N_{\max}$ cutoff on the convergence of $B(E2)$ (Fig.~\ref{NmaxConvergence}a and b) shows that, for both model spaces $\mathfrak{C}$-1 and $\mathfrak{C}$-2, large $N_{\max}$ values are required in order to reach convergence. Indeed, we find that, while convergence for the $gs$ rotational band is achieved around $N_{\max}=12$, the Hoyle-state rotational band requires at least $N_{\max}=18$ for convergence. Similar dependence on $N_{\max}$ is found for the matter rms radius of the ground state and for the electric quadrupole moment (Fig.~\ref{NmaxConvergence}c and d, respectively), both of which require at least $N_{\max}=12$ for convergence. The dependence on $N_{\max}$ does not improve with inclusion of additional symplectic irreps; that is, convergence cannot be achieved with low $N_{\max}$ and many symplectic irreps. These observations underscore the importance of high $N_{\max}$ values for achieving converged $B(E2)$ strengths. Such $N_{\max}$ values are within reach of the NCSpM but well-beyond that of the NCSM calculations due to the combinatorial growth of its model spaces with increasing $N_{\max}$ values.
\begin{figure*}[th]
\centering
(a) \hspace{3in} (b)\\
\includegraphics[width=0.49\textwidth]{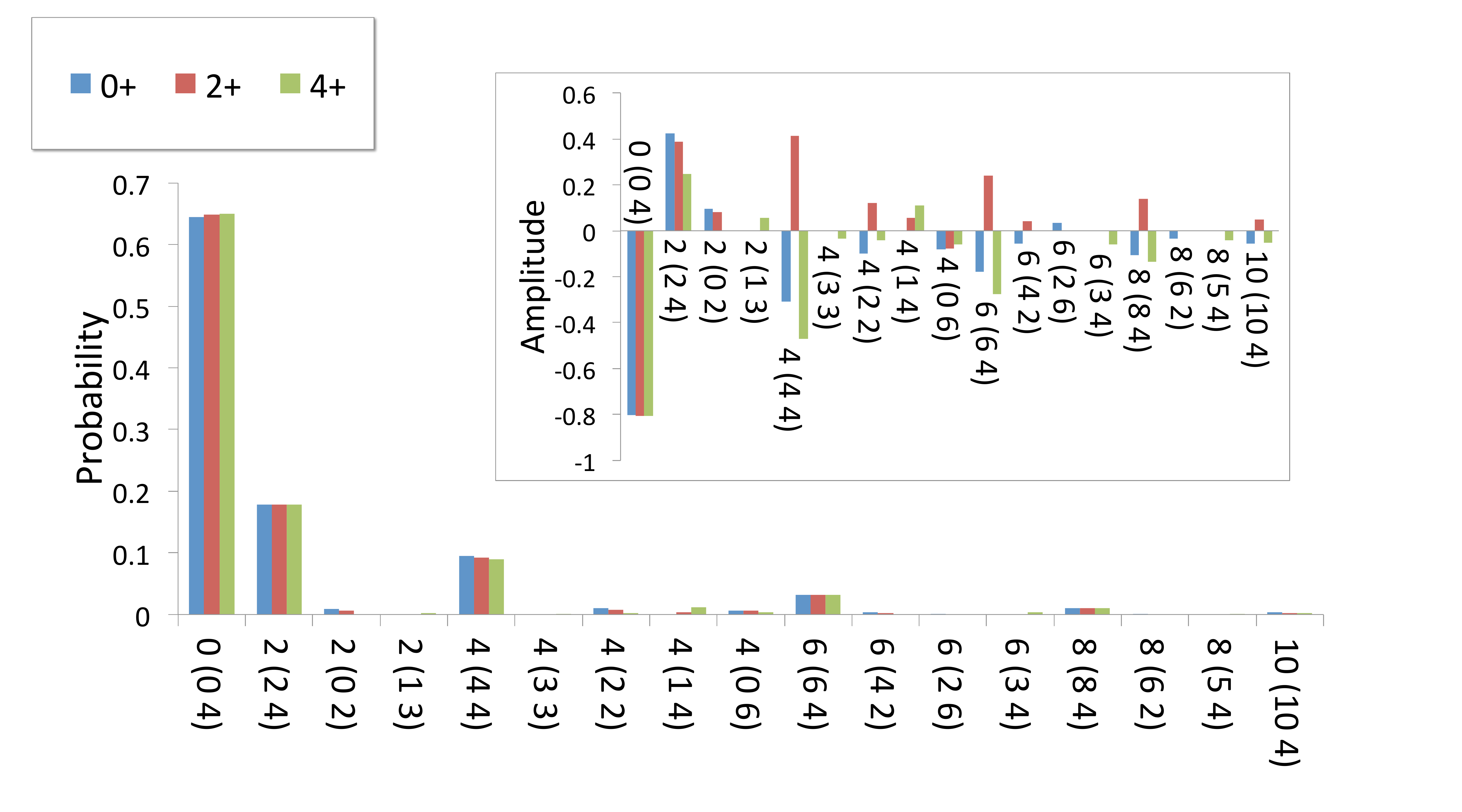}
\includegraphics[width=0.49\textwidth]{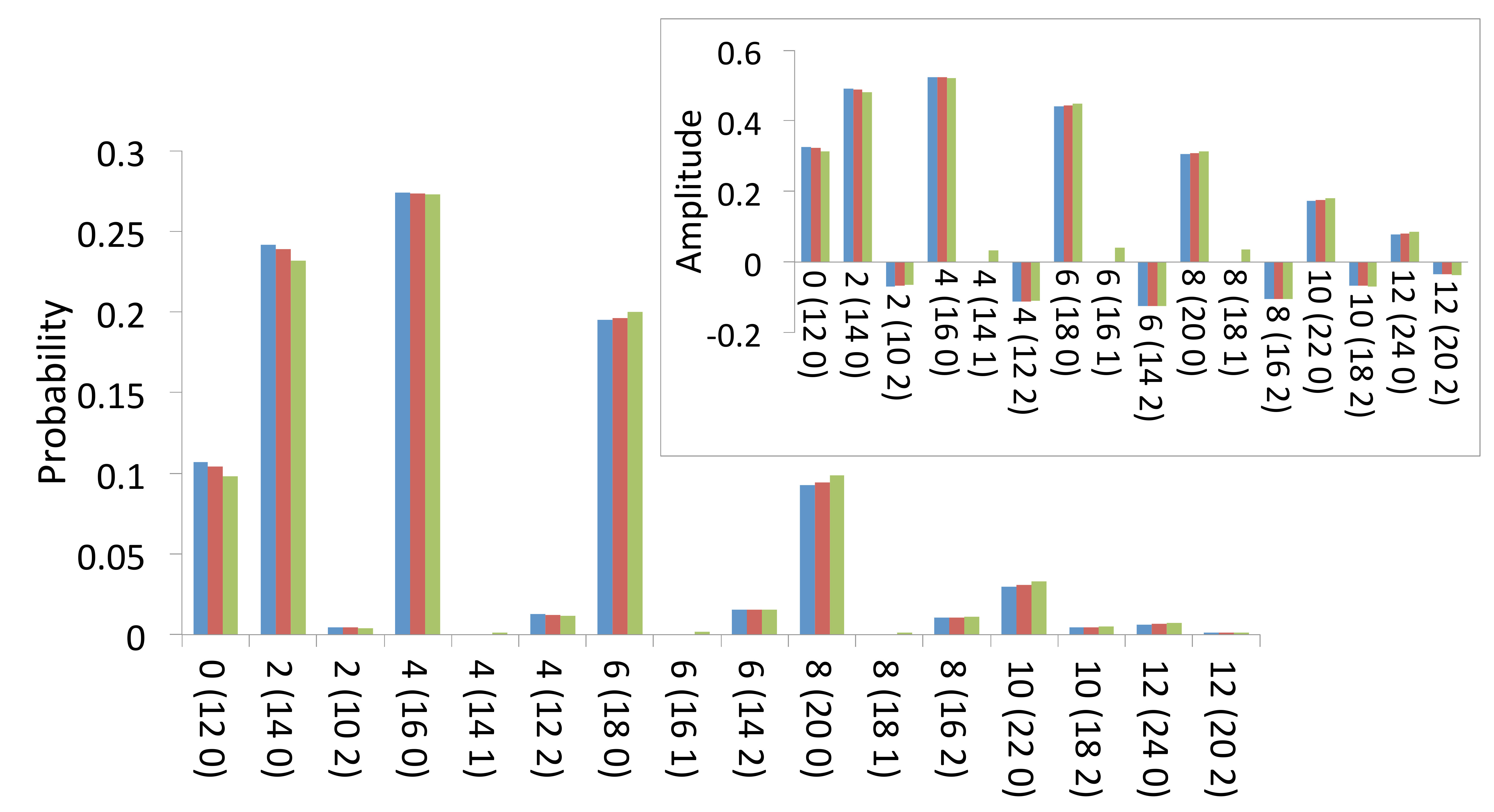}\\
(c)\\
\includegraphics[width=0.49\textwidth]{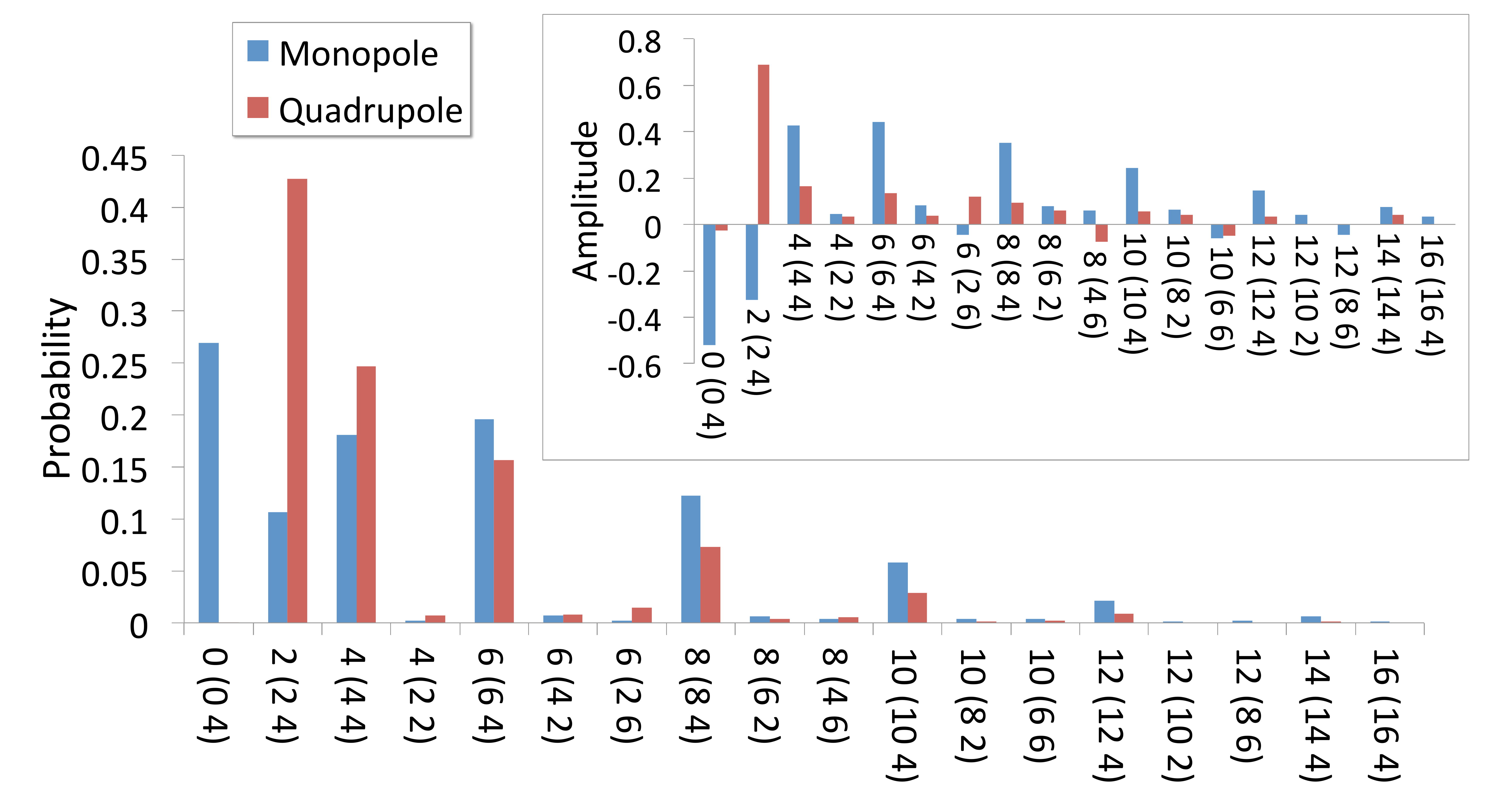}
  \caption{(Color online) NCSpM probabilities and amplitudes (insets) for (a) the ground-state rotational band, (b) the Hoyle-state rotational band, and (c) the GMR and GQR for $^{12}$C. States with probabilities $\geq0.1\%$, which make up $98.83\%-99.63\%$ of the wave functions, are included in the figures.}
  \label{fig:wavfns}
\end{figure*}

\noindent\textbf{Dependence on model parameters} -- The strength parameter $\gamma$ effectively determines to what extent higher-order many-body interactions will contribute to the calculation. A study of its effect on the $^{12}$C energy spectrum (Fig.~\ref{12C_gammaspanNmax20}a) reveals that the additional degree of freedom associated with the $\gamma$ model parameter is substantially limited by the lowest $0^+$ states (with only a small effect on the $gs$ rotational band). Indeed, given the dramatic variation with $\gamma$ for the $0_2^+$ and $0_3^+$ levels, there is only a small range of reasonable $\gamma$ values. In this range, energies and other observables, such as rms matter radii, $B(E2)$ transition rates, and the electric quadrupole moment (see Fig.~\ref{12C_gammaspanNmax20}b-d, respectively),  are found to be in agreement with experiment. 
As the $\gamma$ value decreases from the value adopted in this model (with a limit $\gamma\rightarrow0$, for which the NCSpM simplifies to a multi-shell Elliott model), higher-shell excitations become energetically more favorable and the nucleus expands spatially. This is accompanied by enhancement of collectivity and by considerably larger $B(E2)$ transition strengths. Hence, the second and third $0^+$ states of large deformation fall below the $0^+_{gs}$ state for small values of $\gamma$ (Fig.~\ref{12C_gammaspanNmax20}a). In the limit $\gamma\rightarrow\infty$, the Hamiltonian becomes a harmonic-oscillator potential plus a spin-orbit force. In this case, lowest-energy configurations are favored, and the energy of the \ph{2} state is about 2\ho~MeV lower than that  of the \ph{4} state. It is then remarkable that for the value of the $\gamma$ parameter adopted in this study -- which yields reasonable reproduction of the Hoyle state --  energy spectra and other observables in $p$- and $sd$-shell nuclei are found in a reasonable agreement with their experimental counterparts without further adjustment \cite{LauneyDDTFLDMVB13,TobinFLDDB14}.

The spin-orbit strength $\kappa$ is selected using an empirical estimate (see Sec.~\ref{Interaction}), and is not adjusted in the present calculations. However, a $\pm20\%$ variation of the $\kappa$ parameter shows changes of less than $\pm1$ MeV for states in the low-lying energy spectrum (see inset of Fig. 4a in Ref.~\cite{DreyfussLDDB13}), and has no considerable effect on the other observables under consideration ($0.05\%$ to $3\%$).

\subsection{Deformation and giant resonances \label{GR}}
Important information about deformation is found through analysis of the SU(3) $(\lambda\,\mu)$ configurations that comprise the NCSpM wave function. This is based on an established mapping \cite{RosensteelR77b,LeschberD87,CastanosDL88} between the SU(3) $(\lambda\,\mu)$ labels and the shape variables used in the Bohr-Mottelson collective model \cite{BohrMottelson69}. In particular, for large deformation, the labels $(\lambda\,0)$ and $(0\,\mu)$ can be associated with distinctly prolate and oblate shapes, respectively. From this, it is clear that, while the predominant component of the lowest $0^+$ state in $^{12}$C is at 0\ho~and manifests an evident oblate shape (Fig.~\ref{fig:wavfns}a), 
\renewcommand{\arraystretch}{1.2}
\setlength{\tabcolsep}{5pt}
\begin{table}[h]
\centering
\begin{tabular}{ccccc}
					&	$E_{\mathrm{GMR}}$ 		&	$B(E2; \uparrow)$	&	$E_{\mathrm{GQR}}$ 		&	$B(E2; \downarrow)$	\\
					&	(MeV) 			&	 (W.u.)			&	(MeV)			&	(W.u.)				\\ \hline \hline \noalign{\smallskip}
$^{12}$C				&	27.90			& 	2.38				&	20.87			&	7.43					\\ \hline \noalign{\smallskip}
$^{16}$O				&	29.35			& 	21.94			&	23.54			&	8.13					\\ \hline \noalign{\smallskip}
$^{20}$O				&	23.61			& 	6.82				&	23.40			&	3.58					\\ \hline \noalign{\smallskip}
$^{20}$Mg			&	23.61			& 	15.35			&	23.40			&	8.05					\\ \hline \noalign{\smallskip}
$^{20}$Ne				&	24.27			& 	11.94			&	24.39			&	5.90					\\ \hline \noalign{\smallskip}
$^{22}$Mg			&	25.17			& 	13.16			&	24.97			&	6.43					\\ \hline \noalign{\smallskip}
$^{22}$Ne				&	25.17			& 	9.14				&	24.97			&	4.46					\\ \hline \noalign{\smallskip}
\end{tabular}
\caption{\label{tab:resonances}Energies in MeV of the  first excited $0^{+}$ state, $E_{\mathrm{GMR}}$, and the lowest excited $2^{+}$ state that peaks above $0\ho$, $E_{\mathrm{GQR}}$, within the ground-state symplectic irrep for selected $p$- and $sd$-shell nuclei, and their associated $B(E2)$ transition rates in W.u., $B(E2; \uparrow)$ for $0^{+}_{\mathrm{GMR}}\rightarrow 2^{+}_{1}$ and $B(E2; \downarrow)$ for $2^{+}_{\mathrm{GQR}}\rightarrow 0^{+}_{gs}$, calculated with the NCSpM using model space $\mathfrak{C}$-1.}
\end{table}
the second $0^+$ state (Hoyle state) peaks around 8\ho~with a clear indication of a prolate shape deformation, with $(16\,0)$ being the largest  contribution (Fig.~\ref{fig:wavfns}b). The strong prolate deformation of this $0^+_2$ state together with the significance of the \ph{4} symplectic irrep (built on a configuration of three alpha particles, each occupying a single HO shell) indicate that this $0^+$ state has an underlying alpha-particle cluster structure. This points to a need for next-generation NCSM models, which are capable of \emph{ab initio} calculations in larger model spaces, in order to capture important structural information for the Hoyle state. 

In light nuclei, both the GMR and GQR are expected to be broad resonances, of width a few hundred keV, and are particularly difficult to identify experimentally because of their large overlap with other multipolarities (see, e.g., \cite{Brandenberg1985}). The GMR is understood to be the first $0^+$ excitation  of the $gs$ symplectic irrep \cite{BahriDCR90}, which is a breathing mode with a similar shape to that of the ground state (see Fig.~\ref{fig:wavfns}c for $^{12}$C). The GQR candidates are identified as part of the $gs$ symplectic irrep as the  lowest excited $2^{+}$ state that peaks above $0\ho$ (Table \ref{tab:resonances}). For example, for $^{20}$Mg, the $gs$ symplectic irrep adopted is the one that builds upon the most deformed 0\ho~configuration $(4\,2)$ -- for this irrep, the first excited $0^+$ state  has a broad peak with its maximum at 2\ho~with $(6\,2)$ being the most dominant contribution, while the third $2^+$ state exhibits a broad peak with  a dominant  $2\ho(6\,2)$ configuration (note that the two lowest $2^+$ states for this irrep peak at $0\ho$ and are part of the $gs$ rotational band). These dominant configurations represent excitations of the symplectic bandhead induced by the $A^{(2\,0)}_{L}$ symplectic generators with $L=0$ for the GMRs  (or equally, by the monopole operator) and $L=2$ for GQRs  (or equally, by the quadrupole operator). In general, the main contributions to both GMRs and GQRs arise from excitations described by multiples of the $A^{(2\,0)}$  operators. For $^{12}$C, both the GMR and GQR have non-negligible contributions up to $N_{\max}=14$ (Fig.~\ref{fig:wavfns}c). Because the giant resonances are very broad in light nuclei, the inclusion of higher $N_{\max}$ configurations is critical for describing their structure. 

Previous studies of the GQR for $^{16}$O in the symplectic framework identify the resonance near $E_{x}=25$ MeV with a $B(E2; \downarrow)\approx 17\,\mathrm{e}^2\mathrm{fm}^4$ or 10 W.u. \cite{RBVArickx1990}. The NCSpM corroborates these results (Table~\ref{tab:resonances}): it identifies the second $2^+$ excitation of the $gs$ symplectic irrep of $^{16}$O at $23.54$ MeV as having a similar dominant 2\ho~\ph{1} configuration, with a strong $B(E2)$ transition to the ground state. Analysis of the GMR and GQR candidates for a selection of $p$- and $sd$-shell nuclei shows the two resonances close in energy, with a typically higher energy for the breathing mode. Notably, the oblate GMR for $^{12}$C appears much higher in energy than the prolate \ph{4} deformed state near the Hoyle-state energy.

\section{Conclusion}

We carried out a study of the NCSpM in applications to $^{12}$C as well as to giant monopole and quadrupole resonances in light and intermediate-mass nuclei. Previous studies have successfully employed the NCSpM to describe low-lying states of  various $p$- and $sd$-shell nuclei without any parameter adjustment. Here, we show that the NCSpM is capable of describing $\alpha$-clustering in the Hoyle-state rotational band together with the breathing mode in  $^{12}$C, and discuss dependences of the results on the model space considered and on $\gamma$, the only adjustable parameter in the Hamiltonian. We note that the other model parameters are kept fixed: $\ho$ and $\kappa$ are empirically estimated based on the mass $A$, and $\chi$ is selected through self-consistent methods as described in Sec. \ref{Interaction}. 

By varying both the number of symplectic  irreps we include in the model space and the $N_{\max}$ cutoff, we examined the dependence of NCSpM results on the horizontal  and vertical expansion of the model space. We found that including only the spin-0 symplectic irreps built on the most deformed bandheads and extended up to $N_{\rm max}=20$ describes a compressed energy spectrum of $^{12}$C, for which a \ph{4} $0^+$ state with the Hoyle-state properties was found to lie low in energy (between the $2^+$ and $4^+$ states of the $gs$ rotational band). By including one additional spin-1 $0\ho$~symplectic irrep, we showed the importance of the spin-orbit  interaction for reproducing the energy spacing in the $^{12}$C spectrum. However, the inclusion of additional \SpR{3} irreps did not have any effect on the calculations for the $gs$ or Hoyle-state rotational bands, indicating that this model space ($\mathfrak{C}$-2) is sufficient. We demonstrated the necessity for the inclusion of higher-energy excitations in the model space, both for convergence of observables such as the $B(E2)$ transitions, rms matter radii, and electric quadrupole moments, and to describe the wave functions of the Hoyle-state rotational band. Higher $N_{\max}$ configurations were also shown to be key in describing candidates for the GMR and GQR. 

Most importantly, we showed, for the first time, how both collective and cluster-like structures of $^{12}$C, including the Hoyle state and the breathing mode, emerge from a shell-model framework extended to very high $N_{\max}$ values. The ability of the NCSpM to successfully describe the structure of $^{12}$C and other $p$- and $sd$-shell nuclei with only a small number of basis states allows one to study the underlying physics that  would otherwise require ultra-large shell-model spaces.

\section*{Acknowledgments}
We thank D. Rowe, G. Rosensteel, and J. L. Wood, as well as the PetaApps Collaboration, in particular, J. P. Vary, P. Maris, U. Catalyurek, E. Saule, and M. Sosonkina, for useful discussions. This work was supported by the U.S. NSF (OCI-0904874 \& ACI-1516338), the U.S. DOE (DE-SC0005248), SURA (Southeastern Universities Research Association), Louisiana State
University (through its direct support of J.P.D.'s work with SURA), and the Czech Science Foundation under Grant No. 16-16772S. Computing resources for this work were provided through the Blue Waters sustained-petascale computing project, the Louisiana Optical Network Initiative, as well as LSU ({\tt www.hpc.lsu.edu}). ACD acknowledges support by the U.S. NSF (grant 1004822) through the REU Site in Dept. of Physics \& Astronomy at LSU. 

\bibliographystyle{apsrev}
\bibliography{lsu_new}

\begin{thebibliography}{84}
\expandafter\ifx\csname natexlab\endcsname\relax\def\natexlab#1{#1}\fi
\expandafter\ifx\csname bibnamefont\endcsname\relax
  \def\bibnamefont#1{#1}\fi
\expandafter\ifx\csname bibfnamefont\endcsname\relax
  \def\bibfnamefont#1{#1}\fi
\expandafter\ifx\csname citenamefont\endcsname\relax
  \def\citenamefont#1{#1}\fi
\expandafter\ifx\csname url\endcsname\relax
  \def\url#1{\texttt{#1}}\fi
\expandafter\ifx\csname urlprefix\endcsname\relax\def\urlprefix{URL }\fi
\providecommand{\bibinfo}[2]{#2}
\providecommand{\eprint}[2][]{\url{#2}}

\bibitem[{\citenamefont{Ellis and Engeland}(1970)}]{EllisE70}
\bibinfo{author}{\bibfnamefont{P.~J.} \bibnamefont{Ellis}} \bibnamefont{and}
  \bibinfo{author}{\bibfnamefont{T.}~\bibnamefont{Engeland}},
  \bibinfo{journal}{Nucl. Phys. A} \textbf{\bibinfo{volume}{144}},
  \bibinfo{pages}{161} (\bibinfo{year}{1970}).

\bibitem[{\citenamefont{Engeland and Ellis}(1972)}]{EngelandE72}
\bibinfo{author}{\bibfnamefont{T.}~\bibnamefont{Engeland}} \bibnamefont{and}
  \bibinfo{author}{\bibfnamefont{P.~J.} \bibnamefont{Ellis}},
  \bibinfo{journal}{Nucl. Phys. A} \textbf{\bibinfo{volume}{181}},
  \bibinfo{pages}{368} (\bibinfo{year}{1972}).

\bibitem[{\citenamefont{Uegaki et~al.}(1977)\citenamefont{Uegaki, Okabe, Abe,
  and Tanaka}}]{UegakiOAT77}
\bibinfo{author}{\bibfnamefont{E.}~\bibnamefont{Uegaki}},
  \bibinfo{author}{\bibfnamefont{S.}~\bibnamefont{Okabe}},
  \bibinfo{author}{\bibfnamefont{Y.}~\bibnamefont{Abe}}, \bibnamefont{and}
  \bibinfo{author}{\bibfnamefont{H.}~\bibnamefont{Tanaka}},
  \bibinfo{journal}{Prog. Theor. Phys.} \textbf{\bibinfo{volume}{57}},
  \bibinfo{pages}{1262} (\bibinfo{year}{1977}).

\bibitem[{\citenamefont{Kamimura et~al.}(1981)}]{Kamimura81}
\bibinfo{author}{\bibfnamefont{M.}~\bibnamefont{Kamimura}}
  \bibnamefont{et~al.}, \bibinfo{journal}{Nucl. Phys. A}
  \textbf{\bibinfo{volume}{351}}, \bibinfo{pages}{456} (\bibinfo{year}{1981}).

\bibitem[{\citenamefont{Suzuki and Hecht}(1986)}]{SuzukiH86}
\bibinfo{author}{\bibfnamefont{Y.}~\bibnamefont{Suzuki}} \bibnamefont{and}
  \bibinfo{author}{\bibfnamefont{K.~T.} \bibnamefont{Hecht}},
  \bibinfo{journal}{Nucl. Phys. A} \textbf{\bibinfo{volume}{455}},
  \bibinfo{pages}{315} (\bibinfo{year}{1986}).

\bibitem[{\citenamefont{Burbidge et~al.}(1957)\citenamefont{Burbidge, Burbidge,
  Fowler, and Hoyle}}]{B2FH}
\bibinfo{author}{\bibfnamefont{E.~M.} \bibnamefont{Burbidge}},
  \bibinfo{author}{\bibfnamefont{G.~R.} \bibnamefont{Burbidge}},
  \bibinfo{author}{\bibfnamefont{W.~A.} \bibnamefont{Fowler}},
  \bibnamefont{and} \bibinfo{author}{\bibfnamefont{F.}~\bibnamefont{Hoyle}},
  \bibinfo{journal}{Rev. Mod. Phys.} \textbf{\bibinfo{volume}{29}},
  \bibinfo{pages}{547} (\bibinfo{year}{1957}).

\bibitem[{\citenamefont{Kanada-En'yo}(1998)}]{Kanada98}
\bibinfo{author}{\bibfnamefont{Y.}~\bibnamefont{Kanada-En'yo}},
  \bibinfo{journal}{Phys. Rev. Lett.} \textbf{\bibinfo{volume}{81}},
  \bibinfo{pages}{5291} (\bibinfo{year}{1998}).

\bibitem[{\citenamefont{Funaki et~al.}(2003)\citenamefont{Funaki, Tohsaki,
  Horiuchi, Schuck, and Ropke}}]{FunakiTHSR03}
\bibinfo{author}{\bibfnamefont{Y.}~\bibnamefont{Funaki}},
  \bibinfo{author}{\bibfnamefont{A.}~\bibnamefont{Tohsaki}},
  \bibinfo{author}{\bibfnamefont{H.}~\bibnamefont{Horiuchi}},
  \bibinfo{author}{\bibfnamefont{P.}~\bibnamefont{Schuck}}, \bibnamefont{and}
  \bibinfo{author}{\bibfnamefont{G.}~\bibnamefont{Ropke}},
  \bibinfo{journal}{Phys. Rev. C} \textbf{\bibinfo{volume}{67}},
  \bibinfo{pages}{051306} (\bibinfo{year}{2003}).

\bibitem[{\citenamefont{Yamada and Schuck}(2005)}]{YamadaS05}
\bibinfo{author}{\bibfnamefont{T.}~\bibnamefont{Yamada}} \bibnamefont{and}
  \bibinfo{author}{\bibfnamefont{P.}~\bibnamefont{Schuck}},
  \bibinfo{journal}{Eur. Phys. J. A} \textbf{\bibinfo{volume}{26}},
  \bibinfo{pages}{185} (\bibinfo{year}{2005}).

\bibitem[{\citenamefont{Chernykh et~al.}(2007)\citenamefont{Chernykh,
  Feldmeier, Neff, von Neumann-Cosel, and Richter}}]{ChernykhFNNR07}
\bibinfo{author}{\bibfnamefont{M.}~\bibnamefont{Chernykh}},
  \bibinfo{author}{\bibfnamefont{H.}~\bibnamefont{Feldmeier}},
  \bibinfo{author}{\bibfnamefont{T.}~\bibnamefont{Neff}},
  \bibinfo{author}{\bibfnamefont{P.}~\bibnamefont{von Neumann-Cosel}},
  \bibnamefont{and} \bibinfo{author}{\bibfnamefont{A.}~\bibnamefont{Richter}},
  \bibinfo{journal}{Phys. Rev. Lett.} \textbf{\bibinfo{volume}{98}},
  \bibinfo{pages}{032501} (\bibinfo{year}{2007}).

\bibitem[{\citenamefont{Umar et~al.}(2010)\citenamefont{Umar, Maruhn, Itagaki,
  and Oberacker}}]{UmarMIO10}
\bibinfo{author}{\bibfnamefont{A.~S.} \bibnamefont{Umar}},
  \bibinfo{author}{\bibfnamefont{J.~A.} \bibnamefont{Maruhn}},
  \bibinfo{author}{\bibfnamefont{N.}~\bibnamefont{Itagaki}}, \bibnamefont{and}
  \bibinfo{author}{\bibfnamefont{V.~E.} \bibnamefont{Oberacker}},
  \bibinfo{journal}{Phys. Rev. Lett.} \textbf{\bibinfo{volume}{104}},
  \bibinfo{pages}{212503} (\bibinfo{year}{2010}).

\bibitem[{\citenamefont{Khoa et~al.}(2011)\citenamefont{Khoa, Cuonga, and
  Kanada-En'yo}}]{KhoaCK11}
\bibinfo{author}{\bibfnamefont{D.~T.} \bibnamefont{Khoa}},
  \bibinfo{author}{\bibfnamefont{D.~C.} \bibnamefont{Cuonga}},
  \bibnamefont{and}
  \bibinfo{author}{\bibfnamefont{Y.}~\bibnamefont{Kanada-En'yo}},
  \bibinfo{journal}{Phys. Lett. B} \textbf{\bibinfo{volume}{695}},
  \bibinfo{pages}{469} (\bibinfo{year}{2011}).

\bibitem[{\citenamefont{Neff and Feldmeier}(2014)}]{NeffF14}
\bibinfo{author}{\bibfnamefont{T.}~\bibnamefont{Neff}} \bibnamefont{and}
  \bibinfo{author}{\bibfnamefont{H.}~\bibnamefont{Feldmeier}},
  \bibinfo{journal}{J. Phys.: Conf. Ser.} \textbf{\bibinfo{volume}{569}},
  \bibinfo{pages}{012062} (\bibinfo{year}{2014}).

\bibitem[{\citenamefont{Horiuchi et~al.}(2012)\citenamefont{Horiuchi, Ikeda,
  and Kato}}]{HoriuchiIK12}
\bibinfo{author}{\bibfnamefont{H.}~\bibnamefont{Horiuchi}},
  \bibinfo{author}{\bibfnamefont{K.}~\bibnamefont{Ikeda}}, \bibnamefont{and}
  \bibinfo{author}{\bibfnamefont{K.}~\bibnamefont{Kato}},
  \bibinfo{journal}{Prog. Theor. Phys. Suppl.} \textbf{\bibinfo{volume}{192}},
  \bibinfo{pages}{1} (\bibinfo{year}{2012}).

\bibitem[{\citenamefont{Y.~Funaki}(2015)}]{FunakiHT15}
\bibinfo{author}{\bibfnamefont{A.~T.} \bibnamefont{Y.~Funaki},
  \bibfnamefont{H.~Horiuchi}}, \bibinfo{journal}{Prog. Part. Nucl. Phys.}
  \textbf{\bibinfo{volume}{82}}, \bibinfo{pages}{78} (\bibinfo{year}{2015}).

\bibitem[{\citenamefont{Freer et~al.}(2009)}]{Freer0709}
\bibinfo{author}{\bibfnamefont{M.}~\bibnamefont{Freer}} \bibnamefont{et~al.},
  \bibinfo{journal}{Phys. Rev. C} \textbf{\bibinfo{volume}{80}},
  \bibinfo{pages}{041303} (\bibinfo{year}{2009}).

\bibitem[{\citenamefont{Hyldegaard et~al.}(2010)}]{Hyldegaard10}
\bibinfo{author}{\bibfnamefont{S.}~\bibnamefont{Hyldegaard}}
  \bibnamefont{et~al.}, \bibinfo{journal}{Phys. Rev. C}
  \textbf{\bibinfo{volume}{81}}, \bibinfo{pages}{024303}
  (\bibinfo{year}{2010}).

\bibitem[{\citenamefont{Itoh et~al.}(2011)}]{Itoh11}
\bibinfo{author}{\bibfnamefont{M.}~\bibnamefont{Itoh}} \bibnamefont{et~al.},
  \bibinfo{journal}{Phys. Rev. C} \textbf{\bibinfo{volume}{84}},
  \bibinfo{pages}{054308} (\bibinfo{year}{2011}).

\bibitem[{\citenamefont{Zimmerman et~al.}(2011)\citenamefont{Zimmerman,
  Destefano, Freer, Gai, and Smit}}]{ZimmermanDFGS11}
\bibinfo{author}{\bibfnamefont{W.~R.} \bibnamefont{Zimmerman}},
  \bibinfo{author}{\bibfnamefont{N.~E.} \bibnamefont{Destefano}},
  \bibinfo{author}{\bibfnamefont{M.}~\bibnamefont{Freer}},
  \bibinfo{author}{\bibfnamefont{M.}~\bibnamefont{Gai}}, \bibnamefont{and}
  \bibinfo{author}{\bibfnamefont{F.~D.} \bibnamefont{Smit}},
  \bibinfo{journal}{Phys. Rev. C} \textbf{\bibinfo{volume}{84}},
  \bibinfo{pages}{027304} (\bibinfo{year}{2011}).

\bibitem[{\citenamefont{Zimmerman et~al.}(2013)}]{Zimmerman13}
\bibinfo{author}{\bibfnamefont{W.~R.} \bibnamefont{Zimmerman}}
  \bibnamefont{et~al.}, \bibinfo{journal}{Phys. Rev. Lett.}
  \textbf{\bibinfo{volume}{110}}, \bibinfo{pages}{152502}
  (\bibinfo{year}{2013}).

\bibitem[{\citenamefont{Patel}(2013)}]{NPatel13}
\bibinfo{author}{\bibfnamefont{N.~R.} \bibnamefont{Patel}},
  \bibinfo{journal}{PhD Thesis}  (\bibinfo{year}{2013}).

\bibitem[{\citenamefont{Marin-Lambarri
  et~al.}(2014)\citenamefont{Marin-Lambarri, Bijker, Freer, Gai, Kokalova,
  Parker, and Wheldon}}]{Marin14}
\bibinfo{author}{\bibfnamefont{D.~J.} \bibnamefont{Marin-Lambarri}},
  \bibinfo{author}{\bibfnamefont{R.}~\bibnamefont{Bijker}},
  \bibinfo{author}{\bibfnamefont{M.}~\bibnamefont{Freer}},
  \bibinfo{author}{\bibfnamefont{M.}~\bibnamefont{Gai}},
  \bibinfo{author}{\bibfnamefont{T.}~\bibnamefont{Kokalova}},
  \bibinfo{author}{\bibfnamefont{D.~J.} \bibnamefont{Parker}},
  \bibnamefont{and} \bibinfo{author}{\bibfnamefont{C.}~\bibnamefont{Wheldon}},
  \bibinfo{journal}{Phys. Rev. Lett.} \textbf{\bibinfo{volume}{113}},
  \bibinfo{pages}{012502} (\bibinfo{year}{2014}).

\bibitem[{\citenamefont{Freer et~al.}(2010)}]{Freer2010}
\bibinfo{author}{\bibfnamefont{M.}~\bibnamefont{Freer}} \bibnamefont{et~al.},
  \bibinfo{journal}{Nucl. Phys. A} \textbf{\bibinfo{volume}{834}},
  \bibinfo{pages}{621c} (\bibinfo{year}{2010}).

\bibitem[{\citenamefont{Epelbaum et~al.}(2011)\citenamefont{Epelbaum, Krebs,
  Lee, and Meissner}}]{EpelbaumKLM11}
\bibinfo{author}{\bibfnamefont{E.}~\bibnamefont{Epelbaum}},
  \bibinfo{author}{\bibfnamefont{H.}~\bibnamefont{Krebs}},
  \bibinfo{author}{\bibfnamefont{D.}~\bibnamefont{Lee}}, \bibnamefont{and}
  \bibinfo{author}{\bibfnamefont{U.-G.} \bibnamefont{Meissner}},
  \bibinfo{journal}{Phys. Rev. Lett.} \textbf{\bibinfo{volume}{106}},
  \bibinfo{pages}{192501} (\bibinfo{year}{2011}).

\bibitem[{\citenamefont{Epelbaum et~al.}(2012)\citenamefont{Epelbaum, Krebs,
  L{\"a}hde, Lee, and Mei{\ss}ner}}]{EpelbaumKLLM12}
\bibinfo{author}{\bibfnamefont{E.}~\bibnamefont{Epelbaum}},
  \bibinfo{author}{\bibfnamefont{H.}~\bibnamefont{Krebs}},
  \bibinfo{author}{\bibfnamefont{T.}~\bibnamefont{L{\"a}hde}},
  \bibinfo{author}{\bibfnamefont{D.}~\bibnamefont{Lee}}, \bibnamefont{and}
  \bibinfo{author}{\bibfnamefont{U.-G.} \bibnamefont{Mei{\ss}ner}},
  \bibinfo{journal}{Phys. Rev. Lett.} \textbf{\bibinfo{volume}{109}},
  \bibinfo{pages}{252501} (\bibinfo{year}{2012}).

\bibitem[{\citenamefont{Wiringa}(2012)}]{Wiringa12}
\bibinfo{author}{\bibfnamefont{R.~B.} \bibnamefont{Wiringa}},
  \emph{\bibinfo{title}{Wilhelm und Else Heraeus Seminar on Nuclear
  Ground-State Properties of the Lightest Nuclei: Status and Perspectives}}
  (\bibinfo{year}{2012}).

\bibitem[{\citenamefont{Dreyfuss et~al.}(2013)\citenamefont{Dreyfuss, Launey,
  Dytrych, Draayer, and Bahri}}]{DreyfussLDDB13}
\bibinfo{author}{\bibfnamefont{A.~C.} \bibnamefont{Dreyfuss}},
  \bibinfo{author}{\bibfnamefont{K.~D.} \bibnamefont{Launey}},
  \bibinfo{author}{\bibfnamefont{T.}~\bibnamefont{Dytrych}},
  \bibinfo{author}{\bibfnamefont{J.~P.} \bibnamefont{Draayer}},
  \bibnamefont{and} \bibinfo{author}{\bibfnamefont{C.}~\bibnamefont{Bahri}},
  \bibinfo{journal}{Phys. Lett. B} \textbf{\bibinfo{volume}{727}},
  \bibinfo{pages}{511} (\bibinfo{year}{2013}).

\bibitem[{\citenamefont{Fynbo et~al.}(2005)}]{Fynbo05}
\bibinfo{author}{\bibfnamefont{H.~O.~U.} \bibnamefont{Fynbo}}
  \bibnamefont{et~al.}, \bibinfo{journal}{Nature}
  \textbf{\bibinfo{volume}{433}}, \bibinfo{pages}{136} (\bibinfo{year}{2005}).

\bibitem[{\citenamefont{Rowe et~al.}(2006)\citenamefont{Rowe, Thiamova, and
  Wood}}]{RoweTW06}
\bibinfo{author}{\bibfnamefont{D.~J.} \bibnamefont{Rowe}},
  \bibinfo{author}{\bibfnamefont{G.}~\bibnamefont{Thiamova}}, \bibnamefont{and}
  \bibinfo{author}{\bibfnamefont{J.~L.} \bibnamefont{Wood}},
  \bibinfo{journal}{Phys. Rev. Lett.} \textbf{\bibinfo{volume}{97}},
  \bibinfo{pages}{202501} (\bibinfo{year}{2006}).

\bibitem[{\citenamefont{Heyde and Wood}(2011)}]{HeydeW11}
\bibinfo{author}{\bibfnamefont{K.}~\bibnamefont{Heyde}} \bibnamefont{and}
  \bibinfo{author}{\bibfnamefont{J.}~\bibnamefont{Wood}},
  \bibinfo{journal}{Rev. Mod. Phys.} \textbf{\bibinfo{volume}{83}},
  \bibinfo{pages}{1467} (\bibinfo{year}{2011}).

\bibitem[{\citenamefont{Kulp et~al.}(2008)}]{Kulp08}
\bibinfo{author}{\bibfnamefont{W.}~\bibnamefont{Kulp}} \bibnamefont{et~al.},
  \bibinfo{journal}{Phys. Rev. C} \textbf{\bibinfo{volume}{77}},
  \bibinfo{pages}{061301 (R)} (\bibinfo{year}{2008}).

\bibitem[{\citenamefont{Rowe and Wood}(2010)}]{RoweW2010book}
\bibinfo{author}{\bibfnamefont{D.~J.} \bibnamefont{Rowe}} \bibnamefont{and}
  \bibinfo{author}{\bibfnamefont{J.~L.} \bibnamefont{Wood}},
  \emph{\bibinfo{title}{{ Fundamentals of nuclear models: foundational
  models}}} (\bibinfo{publisher}{World Scientific, Singapore},
  \bibinfo{year}{2010}).

\bibitem[{\citenamefont{Launey et~al.}(2013{\natexlab{a}})\citenamefont{Launey,
  Dytrych, Draayer, Tobin, Ferriss, Langr, Dreyfuss, Maris, Vary, and
  Bahri}}]{LauneyDDTFLDMVB13}
\bibinfo{author}{\bibfnamefont{K.~D.} \bibnamefont{Launey}},
  \bibinfo{author}{\bibfnamefont{T.}~\bibnamefont{Dytrych}},
  \bibinfo{author}{\bibfnamefont{J.~P.} \bibnamefont{Draayer}},
  \bibinfo{author}{\bibfnamefont{G.~K.} \bibnamefont{Tobin}},
  \bibinfo{author}{\bibfnamefont{M.~C.} \bibnamefont{Ferriss}},
  \bibinfo{author}{\bibfnamefont{D.}~\bibnamefont{Langr}},
  \bibinfo{author}{\bibfnamefont{A.~C.} \bibnamefont{Dreyfuss}},
  \bibinfo{author}{\bibfnamefont{P.}~\bibnamefont{Maris}},
  \bibinfo{author}{\bibfnamefont{J.~P.} \bibnamefont{Vary}}, \bibnamefont{and}
  \bibinfo{author}{\bibfnamefont{C.}~\bibnamefont{Bahri}}, in
  \emph{\bibinfo{booktitle}{Proceedings of the 5th International Conference on
  ``Fission and properties of neutron-rich nuclei'', ICFN5, November 4 - 10,
  2012, Sanibel Island, Florida}}, edited by
  \bibinfo{editor}{\bibfnamefont{J.~H.} \bibnamefont{Hamilton}}
  \bibnamefont{and} \bibinfo{editor}{\bibfnamefont{A.~V.}
  \bibnamefont{Ramayya}} (\bibinfo{publisher}{World Scientific, Singapore},
  \bibinfo{year}{2013}{\natexlab{a}}), p.~\bibinfo{pages}{29}.

\bibitem[{\citenamefont{Tobin et~al.}(2014)\citenamefont{Tobin, Ferriss,
  Launey, Dytrych, Draayer, and Bahri}}]{TobinFLDDB14}
\bibinfo{author}{\bibfnamefont{G.~K.} \bibnamefont{Tobin}},
  \bibinfo{author}{\bibfnamefont{M.~C.} \bibnamefont{Ferriss}},
  \bibinfo{author}{\bibfnamefont{K.~D.} \bibnamefont{Launey}},
  \bibinfo{author}{\bibfnamefont{T.}~\bibnamefont{Dytrych}},
  \bibinfo{author}{\bibfnamefont{J.~P.} \bibnamefont{Draayer}},
  \bibnamefont{and} \bibinfo{author}{\bibfnamefont{C.}~\bibnamefont{Bahri}},
  \bibinfo{journal}{Phys. Rev. C} \textbf{\bibinfo{volume}{89}},
  \bibinfo{pages}{034312} (\bibinfo{year}{2014}).

\bibitem[{\citenamefont{Walecka}(1962)}]{Walecka1962}
\bibinfo{author}{\bibfnamefont{J.}~\bibnamefont{Walecka}},
  \bibinfo{journal}{Phys. Rev.} \textbf{\bibinfo{volume}{126}},
  \bibinfo{pages}{653} (\bibinfo{year}{1962}).

\bibitem[{\citenamefont{Werntz and Uberall}(1966)}]{UWerntz1966}
\bibinfo{author}{\bibfnamefont{C.}~\bibnamefont{Werntz}} \bibnamefont{and}
  \bibinfo{author}{\bibfnamefont{H.}~\bibnamefont{Uberall}},
  \bibinfo{journal}{Phys. Rev.} \textbf{\bibinfo{volume}{149}},
  \bibinfo{pages}{762} (\bibinfo{year}{1966}).

\bibitem[{\citenamefont{Suzuki}(1987)}]{Suzuki87}
\bibinfo{author}{\bibfnamefont{Y.}~\bibnamefont{Suzuki}},
  \bibinfo{journal}{Nucl. Phys. A} \textbf{\bibinfo{volume}{470}},
  \bibinfo{pages}{119} (\bibinfo{year}{1987}).

\bibitem[{\citenamefont{Suzuki and Hara}(1989)}]{SuzukiH89}
\bibinfo{author}{\bibfnamefont{Y.}~\bibnamefont{Suzuki}} \bibnamefont{and}
  \bibinfo{author}{\bibfnamefont{S.}~\bibnamefont{Hara}},
  \bibinfo{journal}{Phys. Rev. C} \textbf{\bibinfo{volume}{39}},
  \bibinfo{pages}{658} (\bibinfo{year}{1989}).

\bibitem[{\citenamefont{Yamada et~al.}(2012)}]{YamadaFMHIRST12}
\bibinfo{author}{\bibfnamefont{T.}~\bibnamefont{Yamada}} \bibnamefont{et~al.},
  \bibinfo{journal}{Phys. Rev. C} \textbf{\bibinfo{volume}{85}},
  \bibinfo{pages}{034315} (\bibinfo{year}{2012}).

\bibitem[{\citenamefont{Bacca et~al.}(2013)\citenamefont{Bacca, Barnea,
  Leidemann, and Orlandini}}]{BaccaBLO13}
\bibinfo{author}{\bibfnamefont{S.}~\bibnamefont{Bacca}},
  \bibinfo{author}{\bibfnamefont{N.}~\bibnamefont{Barnea}},
  \bibinfo{author}{\bibfnamefont{W.}~\bibnamefont{Leidemann}},
  \bibnamefont{and}
  \bibinfo{author}{\bibfnamefont{G.}~\bibnamefont{Orlandini}},
  \bibinfo{journal}{Phys. Rev. Lett.} \textbf{\bibinfo{volume}{110}},
  \bibinfo{pages}{042503} (\bibinfo{year}{2013}).

\bibitem[{\citenamefont{Bacca et~al.}(2015)\citenamefont{Bacca, Barnea,
  Leidemann, and Orlandini}}]{BaccaBLO15}
\bibinfo{author}{\bibfnamefont{S.}~\bibnamefont{Bacca}},
  \bibinfo{author}{\bibfnamefont{N.}~\bibnamefont{Barnea}},
  \bibinfo{author}{\bibfnamefont{W.}~\bibnamefont{Leidemann}},
  \bibnamefont{and}
  \bibinfo{author}{\bibfnamefont{G.}~\bibnamefont{Orlandini}},
  \bibinfo{journal}{Phys. Rev. C} \textbf{\bibinfo{volume}{91}},
  \bibinfo{pages}{024303} (\bibinfo{year}{2015}).

\bibitem[{\citenamefont{Rosensteel and
  Rowe}(1977{\natexlab{a}})}]{RosensteelR77}
\bibinfo{author}{\bibfnamefont{G.}~\bibnamefont{Rosensteel}} \bibnamefont{and}
  \bibinfo{author}{\bibfnamefont{D.~J.} \bibnamefont{Rowe}},
  \bibinfo{journal}{Phys. Rev. Lett.} \textbf{\bibinfo{volume}{38}},
  \bibinfo{pages}{10} (\bibinfo{year}{1977}{\natexlab{a}}).

\bibitem[{\citenamefont{Rowe}(1985)}]{Rowe85}
\bibinfo{author}{\bibfnamefont{D.~J.} \bibnamefont{Rowe}},
  \bibinfo{journal}{Rep. Progr. Phys.} \textbf{\bibinfo{volume}{48}},
  \bibinfo{pages}{1419} (\bibinfo{year}{1985}).

\bibitem[{\citenamefont{Hecht and Braunschweig}(1978)}]{HechtB82}
\bibinfo{author}{\bibfnamefont{K.~T.} \bibnamefont{Hecht}} \bibnamefont{and}
  \bibinfo{author}{\bibfnamefont{D.}~\bibnamefont{Braunschweig}},
  \bibinfo{journal}{Nucl. Phys. A} \textbf{\bibinfo{volume}{295}},
  \bibinfo{pages}{34} (\bibinfo{year}{1978}).

\bibitem[{\citenamefont{Vassanji and Rowe}(1983)}]{VassanjiR83}
\bibinfo{author}{\bibfnamefont{M.}~\bibnamefont{Vassanji}} \bibnamefont{and}
  \bibinfo{author}{\bibfnamefont{D.}~\bibnamefont{Rowe}},
  \bibinfo{journal}{Phys. Lett. B} \textbf{\bibinfo{volume}{125}},
  \bibinfo{pages}{103} (\bibinfo{year}{1983}).

\bibitem[{\citenamefont{Arickx et~al.}(1990)\citenamefont{Arickx, Broeckhove,
  Vassanji, and Rowe}}]{RBVArickx1990}
\bibinfo{author}{\bibfnamefont{F.}~\bibnamefont{Arickx}},
  \bibinfo{author}{\bibfnamefont{J.}~\bibnamefont{Broeckhove}},
  \bibinfo{author}{\bibfnamefont{M.}~\bibnamefont{Vassanji}}, \bibnamefont{and}
  \bibinfo{author}{\bibfnamefont{D.}~\bibnamefont{Rowe}},
  \bibinfo{journal}{Nucl. Phys. A} \textbf{\bibinfo{volume}{511}},
  \bibinfo{pages}{49} (\bibinfo{year}{1990}).

\bibitem[{\citenamefont{Bahri et~al.}(1990)\citenamefont{Bahri, Draayer,
  Casta{\~n}os, and Rosensteel}}]{BahriDCR90}
\bibinfo{author}{\bibfnamefont{C.}~\bibnamefont{Bahri}},
  \bibinfo{author}{\bibfnamefont{J.~P.} \bibnamefont{Draayer}},
  \bibinfo{author}{\bibfnamefont{O.}~\bibnamefont{Casta{\~n}os}},
  \bibnamefont{and}
  \bibinfo{author}{\bibfnamefont{G.}~\bibnamefont{Rosensteel}},
  \bibinfo{journal}{Phys. Lett. B} \textbf{\bibinfo{volume}{234}},
  \bibinfo{pages}{430} (\bibinfo{year}{1990}).

\bibitem[{\citenamefont{Carvalho and Rowe}(1992)}]{CarvalhoR92}
\bibinfo{author}{\bibfnamefont{J.}~\bibnamefont{Carvalho}} \bibnamefont{and}
  \bibinfo{author}{\bibfnamefont{D.}~\bibnamefont{Rowe}},
  \bibinfo{journal}{Nucl. Phys. A} \textbf{\bibinfo{volume}{548}},
  \bibinfo{pages}{1} (\bibinfo{year}{1992}).

\bibitem[{\citenamefont{Bahri and Rowe}(2000)}]{BahriR00}
\bibinfo{author}{\bibfnamefont{C.}~\bibnamefont{Bahri}} \bibnamefont{and}
  \bibinfo{author}{\bibfnamefont{D.~J.} \bibnamefont{Rowe}},
  \bibinfo{journal}{Nucl. Phys. A} \textbf{\bibinfo{volume}{662}},
  \bibinfo{pages}{125} (\bibinfo{year}{2000}).

\bibitem[{\citenamefont{Carvalho et~al.}(2002)\citenamefont{Carvalho, Rowe,
  Karram, and Bahri}}]{CarvalhoRKB02}
\bibinfo{author}{\bibfnamefont{J.}~\bibnamefont{Carvalho}},
  \bibinfo{author}{\bibfnamefont{D.}~\bibnamefont{Rowe}},
  \bibinfo{author}{\bibfnamefont{S.}~\bibnamefont{Karram}}, \bibnamefont{and}
  \bibinfo{author}{\bibfnamefont{C.}~\bibnamefont{Bahri}},
  \bibinfo{journal}{Nucl. Phys. A} \textbf{\bibinfo{volume}{703}},
  \bibinfo{pages}{167} (\bibinfo{year}{2002}).

\bibitem[{\citenamefont{Dytrych
  et~al.}(2007{\natexlab{a}})\citenamefont{Dytrych, Sviratcheva, Bahri,
  Draayer, and Vary}}]{DytrychSBDV_PRL07}
\bibinfo{author}{\bibfnamefont{T.}~\bibnamefont{Dytrych}},
  \bibinfo{author}{\bibfnamefont{K.~D.} \bibnamefont{Sviratcheva}},
  \bibinfo{author}{\bibfnamefont{C.}~\bibnamefont{Bahri}},
  \bibinfo{author}{\bibfnamefont{J.~P.} \bibnamefont{Draayer}},
  \bibnamefont{and} \bibinfo{author}{\bibfnamefont{J.~P.} \bibnamefont{Vary}},
  \bibinfo{journal}{Phys. Rev. Lett.} \textbf{\bibinfo{volume}{98}},
  \bibinfo{pages}{162503} (\bibinfo{year}{2007}{\natexlab{a}}).

\bibitem[{\citenamefont{Dytrych et~al.}(2008)\citenamefont{Dytrych,
  Sviratcheva, Draayer, Bahri, and Vary}}]{DytrychSDBV08_review}
\bibinfo{author}{\bibfnamefont{T.}~\bibnamefont{Dytrych}},
  \bibinfo{author}{\bibfnamefont{K.~D.} \bibnamefont{Sviratcheva}},
  \bibinfo{author}{\bibfnamefont{J.~P.} \bibnamefont{Draayer}},
  \bibinfo{author}{\bibfnamefont{C.}~\bibnamefont{Bahri}}, \bibnamefont{and}
  \bibinfo{author}{\bibfnamefont{J.~P.} \bibnamefont{Vary}},
  \bibinfo{journal}{J. Phys. G: Nucl. Part. Phys.}
  \textbf{\bibinfo{volume}{35}}, \bibinfo{pages}{123101}
  (\bibinfo{year}{2008}).

\bibitem[{\citenamefont{Dytrych
  et~al.}(2007{\natexlab{b}})\citenamefont{Dytrych, Sviratcheva, Bahri,
  Draayer, and Vary}}]{DytrychSBDV_PRCa07}
\bibinfo{author}{\bibfnamefont{T.}~\bibnamefont{Dytrych}},
  \bibinfo{author}{\bibfnamefont{K.~D.} \bibnamefont{Sviratcheva}},
  \bibinfo{author}{\bibfnamefont{C.}~\bibnamefont{Bahri}},
  \bibinfo{author}{\bibfnamefont{J.~P.} \bibnamefont{Draayer}},
  \bibnamefont{and} \bibinfo{author}{\bibfnamefont{J.~P.} \bibnamefont{Vary}},
  \bibinfo{journal}{Phys. Rev. C} \textbf{\bibinfo{volume}{76}},
  \bibinfo{pages}{014315} (\bibinfo{year}{2007}{\natexlab{b}}).

\bibitem[{\citenamefont{Dytrych et~al.}(2013)\citenamefont{Dytrych, Launey,
  Draayer, Maris, Vary, Saule, Catalyurek, Sosonkina, Langr, and
  Caprio}}]{DytrychLMCDVL_PRL12}
\bibinfo{author}{\bibfnamefont{T.}~\bibnamefont{Dytrych}},
  \bibinfo{author}{\bibfnamefont{K.~D.} \bibnamefont{Launey}},
  \bibinfo{author}{\bibfnamefont{J.~P.} \bibnamefont{Draayer}},
  \bibinfo{author}{\bibfnamefont{P.}~\bibnamefont{Maris}},
  \bibinfo{author}{\bibfnamefont{J.~P.} \bibnamefont{Vary}},
  \bibinfo{author}{\bibfnamefont{E.}~\bibnamefont{Saule}},
  \bibinfo{author}{\bibfnamefont{U.}~\bibnamefont{Catalyurek}},
  \bibinfo{author}{\bibfnamefont{M.}~\bibnamefont{Sosonkina}},
  \bibinfo{author}{\bibfnamefont{D.}~\bibnamefont{Langr}}, \bibnamefont{and}
  \bibinfo{author}{\bibfnamefont{M.~A.} \bibnamefont{Caprio}},
  \bibinfo{journal}{Phys. Rev. Lett.} \textbf{\bibinfo{volume}{111}},
  \bibinfo{pages}{252501} (\bibinfo{year}{2013}).

\bibitem[{\citenamefont{Elliott}(1958{\natexlab{a}})}]{Elliott58}
\bibinfo{author}{\bibfnamefont{J.~P.} \bibnamefont{Elliott}},
  \bibinfo{journal}{Proc. Roy. Soc. A} \textbf{\bibinfo{volume}{245}},
  \bibinfo{pages}{128} (\bibinfo{year}{1958}{\natexlab{a}}).

\bibitem[{\citenamefont{Elliott}(1958{\natexlab{b}})}]{Elliott58b}
\bibinfo{author}{\bibfnamefont{J.~P.} \bibnamefont{Elliott}},
  \bibinfo{journal}{Proc. Roy. Soc. A} \textbf{\bibinfo{volume}{245}},
  \bibinfo{pages}{562} (\bibinfo{year}{1958}{\natexlab{b}}).

\bibitem[{\citenamefont{Elliott and Harvey}(1962)}]{ElliottH62}
\bibinfo{author}{\bibfnamefont{J.~P.} \bibnamefont{Elliott}} \bibnamefont{and}
  \bibinfo{author}{\bibfnamefont{M.}~\bibnamefont{Harvey}},
  \bibinfo{journal}{Proc. Roy. Soc. A} \textbf{\bibinfo{volume}{272}},
  \bibinfo{pages}{557} (\bibinfo{year}{1962}).

\bibitem[{\citenamefont{Navr\'{a}til et~al.}(2000)\citenamefont{Navr\'{a}til,
  Vary, and Barrett}}]{NavratilVB00}
\bibinfo{author}{\bibfnamefont{P.}~\bibnamefont{Navr\'{a}til}},
  \bibinfo{author}{\bibfnamefont{J.~P.} \bibnamefont{Vary}}, \bibnamefont{and}
  \bibinfo{author}{\bibfnamefont{B.~R.} \bibnamefont{Barrett}},
  \bibinfo{journal}{Phys. Rev. Lett.} \textbf{\bibinfo{volume}{84}},
  \bibinfo{pages}{5728} (\bibinfo{year}{2000}).

\bibitem[{\citenamefont{Barrett et~al.}(2013)\citenamefont{Barrett,
  Navr\'{a}til, and Vary}}]{BarrettNV13}
\bibinfo{author}{\bibfnamefont{B.}~\bibnamefont{Barrett}},
  \bibinfo{author}{\bibfnamefont{P.}~\bibnamefont{Navr\'{a}til}},
  \bibnamefont{and} \bibinfo{author}{\bibfnamefont{J.}~\bibnamefont{Vary}},
  \bibinfo{journal}{Prog. Part. Nucl. Phys.} \textbf{\bibinfo{volume}{69}},
  \bibinfo{pages}{131} (\bibinfo{year}{2013}).

\bibitem[{\citenamefont{Rosensteel and
  Rowe}(1977{\natexlab{b}})}]{RosensteelR77b}
\bibinfo{author}{\bibfnamefont{G.}~\bibnamefont{Rosensteel}} \bibnamefont{and}
  \bibinfo{author}{\bibfnamefont{D.~J.} \bibnamefont{Rowe}},
  \bibinfo{journal}{Ann. Phys. N.Y.} \textbf{\bibinfo{volume}{104}},
  \bibinfo{pages}{134} (\bibinfo{year}{1977}{\natexlab{b}}).

\bibitem[{\citenamefont{Leschber and Draayer}(1987)}]{LeschberD87}
\bibinfo{author}{\bibfnamefont{Y.}~\bibnamefont{Leschber}} \bibnamefont{and}
  \bibinfo{author}{\bibfnamefont{J.~P.} \bibnamefont{Draayer}},
  \bibinfo{journal}{Phys. Lett. B} \textbf{\bibinfo{volume}{190}},
  \bibinfo{pages}{1} (\bibinfo{year}{1987}).

\bibitem[{\citenamefont{Casta{\~n}os et~al.}(1988)\citenamefont{Casta{\~n}os,
  Draayer, and Leschber}}]{CastanosDL88}
\bibinfo{author}{\bibfnamefont{O.}~\bibnamefont{Casta{\~n}os}},
  \bibinfo{author}{\bibfnamefont{J.~P.} \bibnamefont{Draayer}},
  \bibnamefont{and} \bibinfo{author}{\bibfnamefont{Y.}~\bibnamefont{Leschber}},
  \bibinfo{journal}{Z. Phys. A} \textbf{\bibinfo{volume}{329}},
  \bibinfo{pages}{33} (\bibinfo{year}{1988}).

\bibitem[{\citenamefont{Draayer et~al.}(1984)\citenamefont{Draayer, Weeks, and
  Rosensteel}}]{DraayerWR84}
\bibinfo{author}{\bibfnamefont{J.}~\bibnamefont{Draayer}},
  \bibinfo{author}{\bibfnamefont{K.}~\bibnamefont{Weeks}}, \bibnamefont{and}
  \bibinfo{author}{\bibfnamefont{G.}~\bibnamefont{Rosensteel}},
  \bibinfo{journal}{Nucl. Phys. A} \textbf{\bibinfo{volume}{419}},
  \bibinfo{pages}{1} (\bibinfo{year}{1984}).

\bibitem[{\citenamefont{Draayer et~al.}(2012)\citenamefont{Draayer, Dytrych,
  Launey, and Langr}}]{DraayerDLL_Erice11}
\bibinfo{author}{\bibfnamefont{J.~P.} \bibnamefont{Draayer}},
  \bibinfo{author}{\bibfnamefont{T.}~\bibnamefont{Dytrych}},
  \bibinfo{author}{\bibfnamefont{K.~D.} \bibnamefont{Launey}},
  \bibnamefont{and} \bibinfo{author}{\bibfnamefont{D.}~\bibnamefont{Langr}},
  \bibinfo{journal}{Prog. Part. Nucl. Phys.} \textbf{\bibinfo{volume}{67}},
  \bibinfo{pages}{515} (\bibinfo{year}{2012}).

\bibitem[{\citenamefont{Launey et~al.}(2016)\citenamefont{Launey, Dytrych, and
  Draayer}}]{REVIEW2016}
\bibinfo{author}{\bibfnamefont{K.~D.} \bibnamefont{Launey}},
  \bibinfo{author}{\bibfnamefont{T.}~\bibnamefont{Dytrych}}, \bibnamefont{and}
  \bibinfo{author}{\bibfnamefont{J.~P.} \bibnamefont{Draayer}},
  \bibinfo{journal}{Prog. Part. Nucl. Phys.} \textbf{\bibinfo{volume}{89}},
  \bibinfo{pages}{101} (\bibinfo{year}{2016}).

\bibitem[{\citenamefont{Dytrych et~al.}(2016)\citenamefont{Dytrych, Maris,
  Launey, Draayer, Vary, Caprio, Langr, Catalyurek, and
  Sosonkina}}]{DytrychMLDVCLCS11}
\bibinfo{author}{\bibfnamefont{T.}~\bibnamefont{Dytrych}},
  \bibinfo{author}{\bibfnamefont{P.}~\bibnamefont{Maris}},
  \bibinfo{author}{\bibfnamefont{K.~D.} \bibnamefont{Launey}},
  \bibinfo{author}{\bibfnamefont{J.~P.} \bibnamefont{Draayer}},
  \bibinfo{author}{\bibfnamefont{J.~P.} \bibnamefont{Vary}},
  \bibinfo{author}{\bibfnamefont{M.}~\bibnamefont{Caprio}},
  \bibinfo{author}{\bibfnamefont{D.}~\bibnamefont{Langr}},
  \bibinfo{author}{\bibfnamefont{U.}~\bibnamefont{Catalyurek}},
  \bibnamefont{and}
  \bibinfo{author}{\bibfnamefont{M.}~\bibnamefont{Sosonkina}},
  \bibinfo{journal}{Comput. Phys. Commun.} \textbf{\bibinfo{volume}{207}}
  (\bibinfo{year}{2016}).

\bibitem[{\citenamefont{Rowe}(2013)}]{Rowe13}
\bibinfo{author}{\bibfnamefont{D.}~\bibnamefont{Rowe}}, \bibinfo{journal}{AIP
  Conf. Proc.} \textbf{\bibinfo{volume}{1541}}, \bibinfo{pages}{104}
  (\bibinfo{year}{2013}).

\bibitem[{\citenamefont{Otsuka et~al.}(2005)\citenamefont{Otsuka, Suzuki,
  Fujimoto, Grawe, and Akaishi}}]{OtsukaSFGA}
\bibinfo{author}{\bibfnamefont{T.}~\bibnamefont{Otsuka}},
  \bibinfo{author}{\bibfnamefont{T.}~\bibnamefont{Suzuki}},
  \bibinfo{author}{\bibfnamefont{R.}~\bibnamefont{Fujimoto}},
  \bibinfo{author}{\bibfnamefont{H.}~\bibnamefont{Grawe}}, \bibnamefont{and}
  \bibinfo{author}{\bibfnamefont{Y.}~\bibnamefont{Akaishi}},
  \bibinfo{journal}{Phys. Rev. Lett.} \textbf{\bibinfo{volume}{95}},
  \bibinfo{pages}{232502} (\bibinfo{year}{2005}).

\bibitem[{\citenamefont{Myo et~al.}(2013)\citenamefont{Myo, Umeya, Toki, and
  Ikeda}}]{MyoUHTI}
\bibinfo{author}{\bibfnamefont{T.}~\bibnamefont{Myo}},
  \bibinfo{author}{\bibfnamefont{A.}~\bibnamefont{Umeya}},
  \bibinfo{author}{\bibfnamefont{H.}~\bibnamefont{Toki}}, \bibnamefont{and}
  \bibinfo{author}{\bibfnamefont{K.}~\bibnamefont{Ikeda}}, \bibinfo{journal}{J.
  Phys.: Conf. Ser.} \textbf{\bibinfo{volume}{436}}, \bibinfo{pages}{012029}
  (\bibinfo{year}{2013}).

\bibitem[{\citenamefont{Harvey}(1968)}]{Harvey68}
\bibinfo{author}{\bibfnamefont{M.}~\bibnamefont{Harvey}},
  \bibinfo{journal}{Adv. Nucl. Phys.} \textbf{\bibinfo{volume}{1}},
  \bibinfo{pages}{67} (\bibinfo{year}{1968}).

\bibitem[{\citenamefont{Rosensteel and Draayer}(1985)}]{RosensteelD85}
\bibinfo{author}{\bibfnamefont{G.}~\bibnamefont{Rosensteel}} \bibnamefont{and}
  \bibinfo{author}{\bibfnamefont{J.~P.} \bibnamefont{Draayer}},
  \bibinfo{journal}{Nucl. Phys. A} \textbf{\bibinfo{volume}{436}},
  \bibinfo{pages}{445} (\bibinfo{year}{1985}).

\bibitem[{\citenamefont{Casta{\~n}os and Draayer}(1989)}]{CastanosD89}
\bibinfo{author}{\bibfnamefont{O.}~\bibnamefont{Casta{\~n}os}}
  \bibnamefont{and} \bibinfo{author}{\bibfnamefont{J.~P.}
  \bibnamefont{Draayer}}, \bibinfo{journal}{Nucl. Phys. A}
  \textbf{\bibinfo{volume}{491}}, \bibinfo{pages}{349 } (\bibinfo{year}{1989}).

\bibitem[{\citenamefont{Rowe}(1967)}]{Rowe67}
\bibinfo{author}{\bibfnamefont{D.~J.} \bibnamefont{Rowe}},
  \bibinfo{journal}{Phys. Rev.} \textbf{\bibinfo{volume}{162}},
  \bibinfo{pages}{866} (\bibinfo{year}{1967}).

\bibitem[{\citenamefont{Bohr and Mottelson}(1969)}]{BohrMottelson69}
\bibinfo{author}{\bibfnamefont{A.}~\bibnamefont{Bohr}} \bibnamefont{and}
  \bibinfo{author}{\bibfnamefont{B.~R.} \bibnamefont{Mottelson}},
  \emph{\bibinfo{title}{Nuclear Structure}}, vol.~\bibinfo{volume}{1}
  (\bibinfo{publisher}{Benjamin, New York}, \bibinfo{year}{1969}).

\bibitem[{\citenamefont{Peterson and Hecht}(1980)}]{PetersonH80}
\bibinfo{author}{\bibfnamefont{D.}~\bibnamefont{Peterson}} \bibnamefont{and}
  \bibinfo{author}{\bibfnamefont{K.}~\bibnamefont{Hecht}},
  \bibinfo{journal}{Nucl. Phys. A} \textbf{\bibinfo{volume}{344}},
  \bibinfo{pages}{361} (\bibinfo{year}{1980}).

\bibitem[{\citenamefont{Shirokov et~al.}(2007)\citenamefont{Shirokov, Vary,
  Mazur, and Weber}}]{ShirokovMZVW07}
\bibinfo{author}{\bibfnamefont{A.}~\bibnamefont{Shirokov}},
  \bibinfo{author}{\bibfnamefont{J.}~\bibnamefont{Vary}},
  \bibinfo{author}{\bibfnamefont{A.}~\bibnamefont{Mazur}}, \bibnamefont{and}
  \bibinfo{author}{\bibfnamefont{T.}~\bibnamefont{Weber}},
  \bibinfo{journal}{Phys. Lett. B} \textbf{\bibinfo{volume}{644}},
  \bibinfo{pages}{33} (\bibinfo{year}{2007}).

\bibitem[{\citenamefont{Ajzenberg-Selove and Kelley}(1990)}]{ASelove90}
\bibinfo{author}{\bibfnamefont{F.}~\bibnamefont{Ajzenberg-Selove}}
  \bibnamefont{and} \bibinfo{author}{\bibfnamefont{J.}~\bibnamefont{Kelley}},
  \bibinfo{journal}{Nucl. Phys. A} \textbf{\bibinfo{volume}{506}},
  \bibinfo{pages}{1} (\bibinfo{year}{1990}).

\bibitem[{\citenamefont{Freer et~al.}(2011)}]{Freer11}
\bibinfo{author}{\bibfnamefont{M.}~\bibnamefont{Freer}} \bibnamefont{et~al.},
  \bibinfo{journal}{Phys. Rev. C} \textbf{\bibinfo{volume}{83}},
  \bibinfo{pages}{034314} (\bibinfo{year}{2011}).

\bibitem[{\citenamefont{Launey et~al.}(2013{\natexlab{b}})\citenamefont{Launey,
  Dreyfuss, Dytrych, Draayer, Langr, Maris, Vary, and
  Bahri}}]{LauneyDDTFLDMVB13b}
\bibinfo{author}{\bibfnamefont{K.~D.} \bibnamefont{Launey}},
  \bibinfo{author}{\bibfnamefont{A.~C.} \bibnamefont{Dreyfuss}},
  \bibinfo{author}{\bibfnamefont{T.}~\bibnamefont{Dytrych}},
  \bibinfo{author}{\bibfnamefont{J.~P.} \bibnamefont{Draayer}},
  \bibinfo{author}{\bibfnamefont{D.}~\bibnamefont{Langr}},
  \bibinfo{author}{\bibfnamefont{P.}~\bibnamefont{Maris}},
  \bibinfo{author}{\bibfnamefont{J.~P.} \bibnamefont{Vary}}, \bibnamefont{and}
  \bibinfo{author}{\bibfnamefont{C.}~\bibnamefont{Bahri}}, \bibinfo{journal}{J.
  Phys.: Conf. Ser.} \textbf{\bibinfo{volume}{436}}, \bibinfo{pages}{012023}
  (\bibinfo{year}{2013}{\natexlab{b}}).

\bibitem[{\citenamefont{Bahri and Draayer}(1994)}]{bahri94rme}
\bibinfo{author}{\bibfnamefont{C.}~\bibnamefont{Bahri}} \bibnamefont{and}
  \bibinfo{author}{\bibfnamefont{J.~P.} \bibnamefont{Draayer}},
  \bibinfo{journal}{Comput. Phys. Commun.} \textbf{\bibinfo{volume}{83}},
  \bibinfo{pages}{59} (\bibinfo{year}{1994}).

\bibitem[{\citenamefont{Draayer and Akiyama}(1973)}]{DraayerSU3_1}
\bibinfo{author}{\bibfnamefont{J.~P.} \bibnamefont{Draayer}} \bibnamefont{and}
  \bibinfo{author}{\bibfnamefont{Y.}~\bibnamefont{Akiyama}},
  \bibinfo{journal}{J. Math. Phys.} \textbf{\bibinfo{volume}{14}},
  \bibinfo{pages}{1904} (\bibinfo{year}{1973}).

\bibitem[{\citenamefont{Morinaga}(1956)}]{PRMorinaga}
\bibinfo{author}{\bibfnamefont{H.}~\bibnamefont{Morinaga}},
  \bibinfo{journal}{Phys. Rev.} \textbf{\bibinfo{volume}{101}},
  \bibinfo{pages}{254} (\bibinfo{year}{1956}).

\bibitem[{\citenamefont{Roth et~al.}(2011)\citenamefont{Roth, Langhammer,
  Calci, Binder, and Navr\'atil}}]{RothLCBN11}
\bibinfo{author}{\bibfnamefont{R.}~\bibnamefont{Roth}},
  \bibinfo{author}{\bibfnamefont{J.}~\bibnamefont{Langhammer}},
  \bibinfo{author}{\bibfnamefont{A.}~\bibnamefont{Calci}},
  \bibinfo{author}{\bibfnamefont{S.}~\bibnamefont{Binder}}, \bibnamefont{and}
  \bibinfo{author}{\bibfnamefont{P.}~\bibnamefont{Navr\'atil}},
  \bibinfo{journal}{Phys. Rev. Lett.} \textbf{\bibinfo{volume}{107}},
  \bibinfo{pages}{072501} (\bibinfo{year}{2011}).

\bibitem[{\citenamefont{Brandenberg}(1985)}]{Brandenberg1985}
\bibinfo{author}{\bibfnamefont{S.}~\bibnamefont{Brandenberg}}, Ph.D. thesis,
  \bibinfo{school}{University of Groningen} (\bibinfo{year}{1985}).

\end{thebibliography}

\end{document}